\documentclass[twocolumn]{aastex631}

\usepackage{latexsym}
\usepackage{natbib}
\usepackage{graphicx}
\usepackage[varg]{txfonts}
\newcommand{\sect}[1]{Sect.\,\ref{#1}}

\newcommand{\app}[1]{Appendix\,\ref{#1}}
\newcommand{\fff}[1]{Fig.\,\ref{#1}}
\newcommand{\figs}[1]{Figs.\,\ref{#1}}
\newcommand{\tab}[1]{Table\,\ref{#1}}

\graphicspath{{./}{figs/}}

\def\widex{85mm}
\def\widcuts{140mm}

\journalinfo{{\small{\rm Accepted for publication in} The Astrophysical Journal, {\rm May 20, 2022}}}

\shorttitle{Parallel plasma loops and the energization of the solar corona}
\shortauthors{Peter et al.}

\begin{document}

%
%
\title{Parallel plasma loops and the energization of the solar corona} 

\correspondingauthor{Hardi Peter}
\email{peter@mps.mpg.de}

\author[0000-0001-9921-0937]{Hardi Peter}
\affiliation{Max Planck Institute for Solar System Research, Justus-von-Liebig-Weg 3, D-37077, G{\" o}ttingen, Germany}

\author[0000-0002-9270-6785]{Lakshmi Pradeep Chitta}
\affiliation{Max Planck Institute for Solar System Research, Justus-von-Liebig-Weg 3, D-37077, G{\" o}ttingen, Germany}

\author[0000-0002-1963-5319]{Feng Chen}
\affiliation{Laboratory for Atmospheric and Space Physics, University of Colorado Boulder, Boulder, CO 80303, USA}

\author[0000-0002-1089-9270]{David I. Pontin}
\affiliation{School of Information and Physical Sciences, University of Newcastle, Callaghan, NSW 2308, Australia}
\affiliation{School of Science and Engineering, University of Dundee, Nethergate, Dundee, DD1 4HN, UK}


\author{Amy R. Winebarger}
\affiliation{NASA/Marshall Space Flight Center, Huntsville, AL 35812, USA}

\author{Leon Golub}
\affiliation{Harvard-Smithsonian Center for Astrophysics, 60 Garden St., Cambridge, MA 02138, USA}

\author[0000-0002-6172-0517]{Sabrina L. Savage}
\affiliation{NASA/Marshall Space Flight Center, Huntsville, AL 35812, USA}

\author{Laurel A. Rachmeler}
\affiliation{National Oceanic and Atmospheric Administration (NOAA), Boulder, CO 80305}

\author{Ken Kobayashi}
\affiliation{NASA/Marshall Space Flight Center, Huntsville, AL 35812, USA}

\author{David H. Brooks}
\affiliation{College of Science, George Mason University, 4400 University Drive, Fairfax, VA 22030, USA}

\author{Jonathan W. Cirtain}
\affiliation{BWX Technologies, Inc., 800 Main St \#400, Lynchburg, VA 24504}

\author{Bart De~Pontieu}
\affiliation{Lockheed Martin Solar and Astrophysics Laboratory, Palo~Alto, CA 94304, USA}
\affiliation{Rosseland Centre for Solar Physics, University of Oslo, P.O. Box 1029 Blindern, NO-0315, Oslo, Norway}
\affiliation{Institute of Theoretical Astrophysics, University of Oslo, P.O. Box 1029 Blindern, NO-0315, Oslo, Norway}

\author{David E. McKenzie}
\affiliation{NASA/Marshall Space Flight Center, Huntsville, AL 35812, USA}

\author{Richard J. Morton}
\affiliation{Department of Mathematics, Physics and Electrical Engineering, Northumbria University, Newcastle Upon Tyne, NE1 8ST, UK}


\author[0000-0002-0405-0668]{Paola Testa}
\affiliation{Harvard-Smithsonian Center for Astrophysics, 60 Garden St., Cambridge, MA 02138, USA}

\author{Sanjiv K. Tiwari}
\affiliation{Bay Area Environmental Research Institute, NASA Research Park, Moffett Field, CA 94035, USA}
\affiliation{Lockheed Martin Solar and Astrophysics Laboratory, Palo~Alto, CA 94304, USA}

\author{Robert W. Walsh}
\affiliation{Jeremiah Horrocks Institute, University of Central Lancashire, Preston, PR1 2HE, UK}

\author{Harry P. Warren}
\affiliation{Space Science Division, Naval Research Laboratory, Washington, DC 20375, USA}

\begin{abstract}
%
%
The outer atmosphere of the Sun is composed of plasma heated to temperatures well in excess of the visible surface.
%
%
We investigate short cool and warm (${<}1$\,MK) loops seen in the core of an active region to address the role of field-line braiding in energising these structures.
%
%
We report observations from the High-resolution Coronal imager (Hi-C) that have been acquired in a coordinated campaign with the Interface Region Imaging Spectrograph (IRIS).
%
%
In the core of the active region, the 172\,\AA\ band of Hi-C and the 1400\,\AA\ channel of IRIS show plasma loops at different temperatures that run in parallel.
There is a small but detectable spatial offset of less than 1{\arcsec} between the loops seen in the two bands.
Most importantly, we do not see observational signatures that these loops might be twisted around each other.
%
%
Considering the scenario of magnetic braiding, our observations of parallel loops imply that the stresses put into the magnetic field have to relax while the braiding is applied: the magnetic field never reaches a highly braided state on these length-scales comparable to the separation of the loops.
This supports recent numerical 3D models of loop braiding in which the effective dissipation is sufficiently large that it keeps the magnetic field from getting highly twisted within a loop.
\end{abstract}
%
%
\keywords{Sun: magnetic fields
      --- Sun: UV radiation 
      --- Sun: transition region
      --- Sun: corona
      --- Magnetohydrodynamics (MHD)} 
%

\section{Introduction\label{S:intro}}

The structure of the upper atmosphere above active regions on the Sun is dominated by coronal loops at temperatures from 0.1 to 10\,MK.
Best seen in extreme UV wavelengths, these loops show heated plasma that is trapped by the magnetic field and therefore  outline magnetic field lines. The energy to heat the plasma and support it against gravity has to be supplied by the magnetic field. Hence, knowing the structure of the magnetic field is a key to understanding the underlying processes that create the hot coronal plasma on the Sun, and also on other stars.
Several concepts to heat the upper atmosphere have been discussed.
Heating by magneto-acoustic waves is considered to be widespread \cite[][]{1983A&A...117..220H,2007Sci...317.1192T}.
However, current observations do not detect sufficient wave energy flux to power the hot parts of active regions   \cite[][]{2011Natur.475..477M}, even though wave motions might hide from detection because of line-of-sight effects or wavelengths and frequencies below the detection limit \cite[][]{2012ApJ...746...31D,2019FrASS...6...38K}.
 An alternative is the energisation of the corona through surface motions resulting in field line braiding \cite[][]{1972ApJ...174..499P,1983ApJ...264..642P,2013Natur.493..501C}, flux-tube tectonics \cite[][]{2002ApJ...576..533P} or injection of plasma, currents or waves \cite[][]{2017Sci...356.1269M,2017ApJ...845L..18D}. %
More recently, also the role of emergence and cancelation of magnetic flux at the footpoints of hot structures has been investigated \cite[][]{2017ApJS..229....4C,2018A&A...615L...9C,2019ApJ...881..107A}.

In a typical active region one can distinguish different types of loops \cite[][]{2014LRSP...11....4R}.
Long warm loops with plasma at temperatures of around 1\,MK have lengths that are a significant fraction of the solar radius and connect to the periphery of an active region (e.g.\ the long loops in \fff{F:hic.full.fov}).
More compact, hot loops, reaching temperatures of up to 5\,MK, are mostly found in the core of active regions \cite[][]{2010ApJ...711..228W}. The core of an active region also hosts an abundance of shorter, cool loops reaching only a few 0.1\,MK,  in particular during phases of emergence of magnetic flux, e.g.\ while an active region forms.\
These cool loops are evident in the 1400\,{\AA} channel of the Interface Region Imaging Spectrograph, IRIS \cite[][]{2014SoPh..289.2733D}.
The emission from the cool loops in this 55\,{\AA} wide wavelength band is clearly dominated by photons from the \ion{Si}{4} doublet at 1393.76\,{\AA} and 1402.77\,{\AA} \cite[][]{2018ApJ...854..174T}, showing that these loops host plasma at temperatures of the order of only 0.1\,MK.
There is also some contribution by the \ion{C}{2} doublet near 1335\,{\AA} which forms at slightly lower temperatures.

Some of these cool loops seen in the IRIS 1400\,{\AA} channel appear without any counterparts visible at higher temperatures around 1\,MK \cite[][]{2014Sci...346C.315P}, 
some have those counterparts \cite[][]{2019A&A...626A..98L}.
%
Still, most observations with the Atmospheric Imaging Assembly, AIA \cite[][]{2012SoPh..275...17L}, in the cores of (emerging) active regions also show short loops in the 171\,{\AA} channel.
Under equilibrium conditions the 171\,{\AA} band has a peak response just below 1\,MK and is dominated by emission from \ion{Fe}{9} at 171.07\,{\AA} (and to a lesser extent by \ion{Fe}{10} at 174.53\,{\AA}).
However, many of the extreme UV channels of AIA also have contributions from lines forming at lower temperatures, e.g., the 171\,{\AA} band has a secondary contribution around 0.3\,MK \cite[][]{2012SoPh..275...41B}.
So the question remains if the short loops in AIA 171\,{\AA} in active region cores are actually originating from temperatures around 0.3\,MK) or higher temperatures (roughly 1\,MK).

Essentially, this raises the question if the loops in the 171\,{\AA} band and the 1400\,{\AA} band are the same structures, or if cool and warm loops co-exist in close vicinity.
More importantly, the relation of the structures seen at 171\,{\AA} and 1400\,{\AA} contain key information on the nature of the cool emerging loops in the active region core.
Comparing observational results to models, we will be able to conclude how the emergence process governs the thermal evolution of the active region core.
We will find that the loops we see in Hi-C 172\,\AA\ and in IRIS 1400\,\AA\ run in parallel, without observable signatures of braiding.
This implies that the magnetic field relaxes already while driven, which is consistent with recent models of field-line braiding.
In principle, it could also be that the loops are aligned with emerging flux emerging too fast to undergo braiding.

In this study we will exploit the potential of the unprecedented spatial resolution of the High-resolution Coronal imager, Hi-C \cite[][]{2019SoPh..294..174R} in the 172\,{\AA} band during its second successful flight (termed Hi-C\,2.1).
Together with data from IRIS and a 3D MHD model of an emerging active region, we can address and settle the above questions.

We will start with a brief discussion of field-line braiding (\sect{S:braid}) before we provide details of the observational data (\sect{S:obs}).
The key observational result of our study, namely that the loops in Hi-C and IRIS run in parallel with a (small) offset, will be presented in \sect{S:results}.
We discuss the temporal evolution and thermal structure of the compact loops in more detail in \sect{sec:tevol} before we compare the observations to one particular 3D MHD model of an emerging loop system in \sect{S:model} and conclude in \sect{S:discussion}.
%

\section{Braiding of coronal structures\label{S:braid}}

Braiding of magnetic field lines by motions at the solar surface will create thin current sheets in the corona, and the dissipated magnetic energy heats the plasma.
This goes along with (bundles of) field lines being braided around each other, as depicted by cartoons in the original papers by \cite{1972ApJ...174..499P,1983ApJ...264..642P}.
Those cartoon representations of the magnetic field would suggest that one should find highly braided structures when observing in the extreme UV, which originates from highly ionized plasma outlining the magnetic field lines.

\begin{figure}
\centerline{\includegraphics[width=85mm]{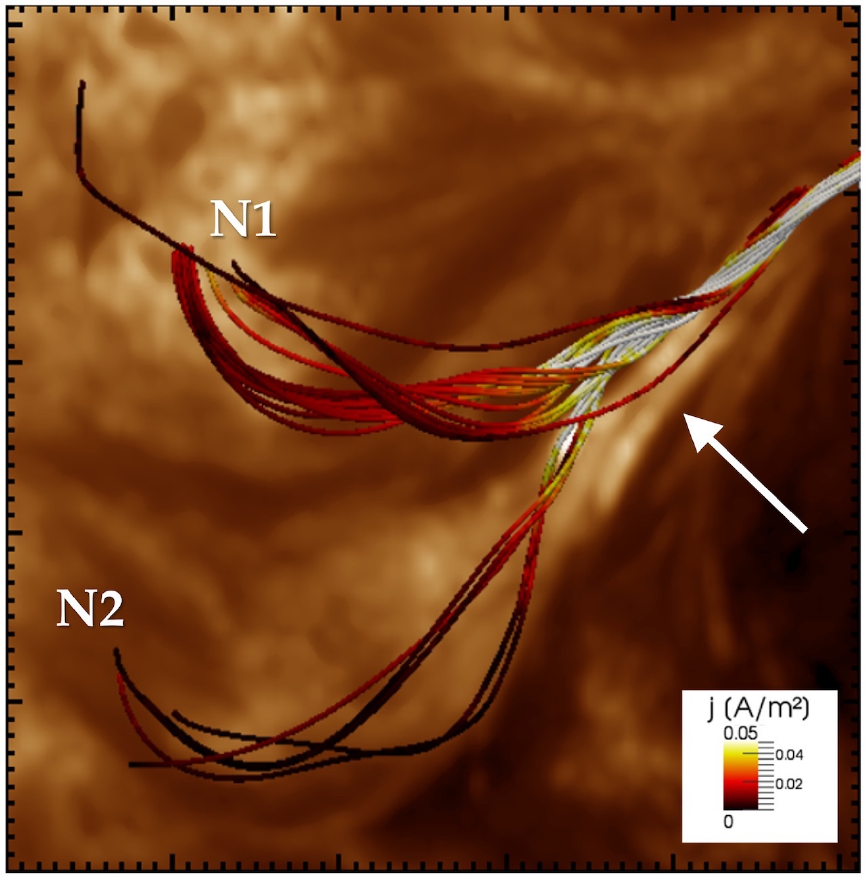}}
\caption{Braiding of a forking flux tube. 
The lines depict the magnetic field of a magnetic flux rope that is forking near the location of the arrow.
Part of the flux rope is rooted near N1, another part near N2.
The background image shows the emission near 193\,\AA\ recorded during the first flight of Hi-C.
The field-of-view is 50$\times$50\,Mm$^2$.
The coloring of the field lines indicates the strength of the currents of the non-linear force-free extrapolation.
The figure is adapted from Fig.\,2c of \cite{2014ApJ...780..102T}.
See \sect{S:braid} for details. 
\label{F:braid.fork}}
\end{figure}

One has to distinguish two cases of braiding of coronal structures:
(1) Near-surface flows could drive braiding within a bundle of field lines that is rooted at each end in one single region in the photosphere, e.g.\ a magnetic flux tube or sheet in the intergranular lanes.
This bundle of field lines would be co-spatial with the loop we see in coronal emission \cite[][]{2012A&A...548A...1P}.
Or in case (2) the loop we see in coronal emission could be hosted by a larger structure with multiple footpoints at one or both ends.
In this case, large-scale shear motions of the multiple footpoints would braid the field and thus inject energy to heat the structure.

In a paper from the first flight of Hi-C, \cite{2013Natur.493..501C} showed the presence of magnetic braiding according to case (2).
They described an elongated feature, resembling a loop, that was stretching above the penumbra of a sunspot.
A detailed study of the magnetic setup of this structure was conducted by \cite{2014ApJ...780..102T} through a non-linear force-free magnetic field extrapolation. 
They found that the main braiding feature of the structure investigated by \cite{2013Natur.493..501C} is located just where a flux-rope forks (arrow in \fff{F:braid.fork}): part of the flux rope is then rooted in N1, part in N2.
Here the separation of the photospheric anchors at this end of the flux rope is comparable to the length of the loop feature.
Probably the shear motions between locations N1 and N2 in \fff{F:braid.fork} contribute to the braiding of the structure. Relating the very same structure as observed by
\cite{2013Natur.493..501C} to the photospheric magnetic field, \cite{2014ApJ...795L..24T} found evidence for flux cancelation. This would be consistent with the interpretation of this structure as a flux rope and would explain why this low-lying structure is heated to high temperatures.

Traditional models for braiding of the magnetic field in coronal loops assume that the magnetic field is braided within the flux tube, i.e.\ according to case (1).
Starting with the models of \cite{1996JGR...10113445G}, there has been an abundance of studies that essentially straighten a coronal loop and put it into a rectangular box.
In these models one can study the braiding in detail \cite[e.g.][]{2008ApJ...677.1348R} and investigate the expected observational signatures of braiding \cite[e.g.][]{2016ApJ...817...47D,2017ApJ...837..108P}.
In these braiding models, either an internal braiding is prescribed \cite[e.g.][]{2010A&A...516A...5W} or it is driven by motions within the modelled bundle of field lines at the footpoints \cite[e.g.][]{2018A&A...615A..84R}.
In \fff{F:braid.david} we depict the magnetic field in a loop braiding model during the relaxation phase.
Here some internal twisting of the field lines is visible.
However, opposed to case (2) exemplified  in \fff{F:braid.fork}, all field lines stay within the envelope of the single loop.
The emission from a coronal loop with a braided magnetic field as shown in \fff{F:braid.david} might not even show any signatures of braiding \cite[][]{2017ApJ...837..108P}.

\begin{figure}
\centerline{\includegraphics[width=85mm]{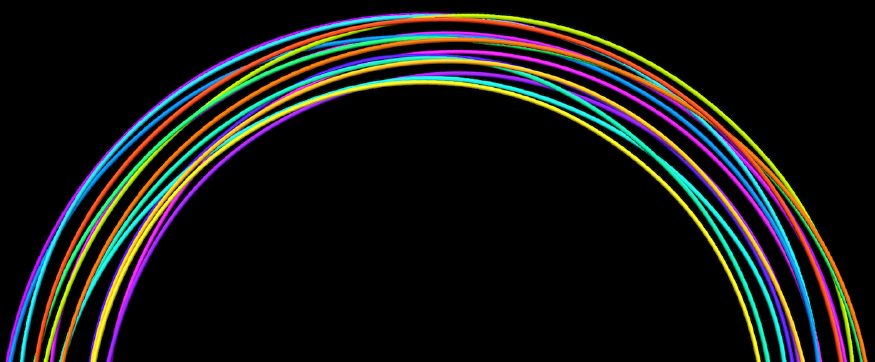}}
\caption{Internal braiding of a bundle of fieldlines in a loop.
The braided magnetic field lines are shown in different colours.
They are are based on a 3D MHD model similar to \cite{2017ApJ...837..108P} where the magnetic field started to relax.
While the original model is for a straight loop, here we curved the appearance to resemble a semi-circular loop.
See \sect{S:braid} for details.
\label{F:braid.david}}
\end{figure}

Still, one could expect that with sufficient spatial resolution one might find braids in the internal structure of a coronal loop.
Numerical 3D MHD models of braiding have been inconclusive in that such internal braids
might or might not be present.
Essentially, this depends on the setup and how quickly the magnetic field
would relax when driven from the footpoints \cite[][]{2016ApJ...817...47D,2017ApJ...837..108P}.

Therefore, new observations will have to show if such braided magnetic fields are observable in the extreme UV, or if the extreme UV images appear to be combed, i.e.\ with structures running parallel.
The main goal of this study is to establish if the short cool and warm loops in the active region core appear to be combed or uncombed.

\begin{figure}
\centerline{\includegraphics[width=85mm]{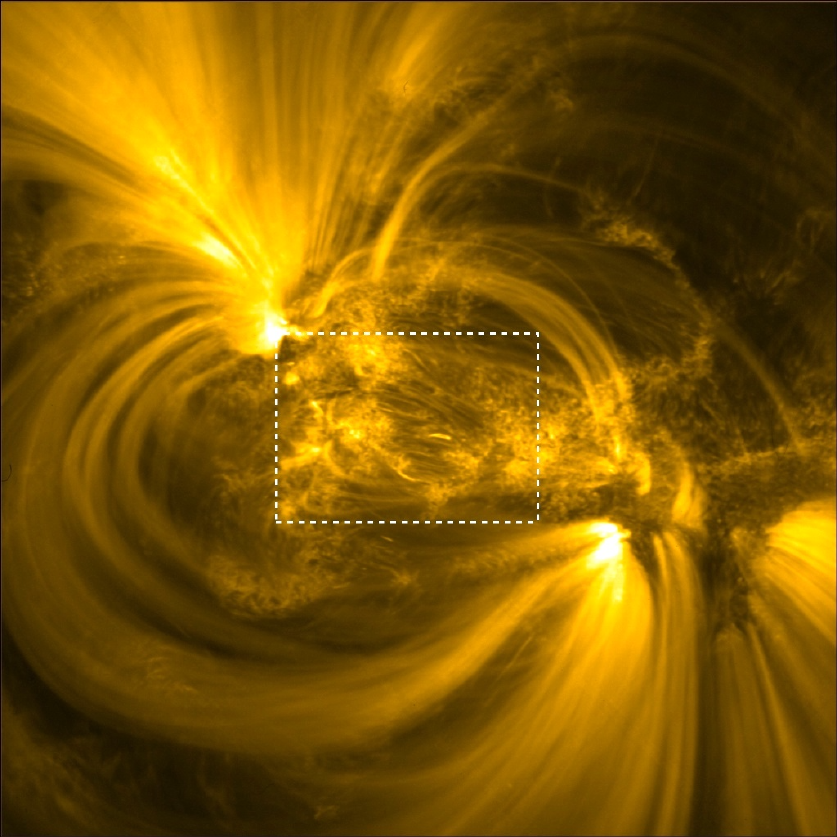}}
\caption{Hi-C image showing warm loops in a solar active region. 
This displays emission around 172\,{\AA} seen by Hi-C over its full field of view of 265{\arcsec} $\times$ 265{\arcsec} at 19:00:33 UT on May 29, 2018.
The center of the image is about 114{\arcsec} east and 259{\arcsec} north of disk center.
The rectangle marks the core of the active region with the abundant small loops.
This 83{\arcsec} $\times$ 60{\arcsec} sub-region is imaged in more detail in \fff{F:hic.core}.
\label{F:hic.full.fov}}
\end{figure}

\section{Observations\label{S:obs}}

We investigate the core of a active region NOAA 12712 as it appeared close to the center of the solar disk.
It showed continuous emergence of new magnetic flux for several days before our observation.
The active region hosts the typical long loops at temperatures of about 1\,MK (see \fff{F:hic.full.fov}) connecting the outer parts of the magnetic concentrations.
In the core of the active region we see numerous shorter loops that connect small patches of opposite magnetic polarity in the region between the sunspots. A zoom into this region is displayed in \fff{F:hic.core}.
Almost the same region of interest has been investigated by \cite{2019ApJ...887...56T} who related small-scale brightenings to the cancellation of magnetic flux in the photosphere below  and its role for heating.  
In our study we will focus on the thermal structure of the compact short loops in the active region core.

\begin{figure*}
\centerline{\includegraphics[width=150mm]{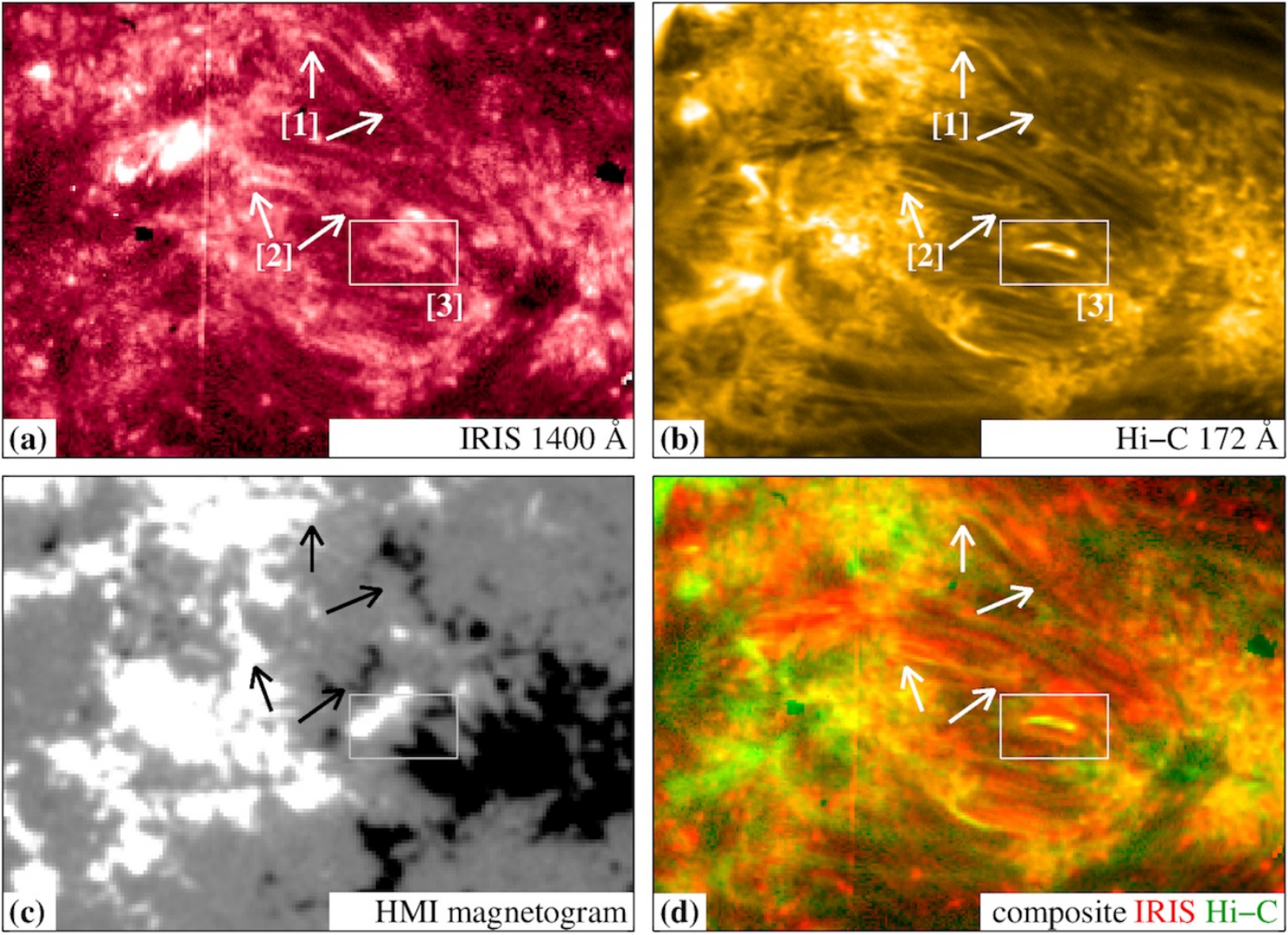}}
\caption{Core of the active region.
The field of view of each panel is 83{\arcsec} $\times$ 60{\arcsec} and corresponds to the rectangle in \fff{F:hic.full.fov}.
Panels (a) and (b) show the images in the IRIS 1400\,{\AA} channel and in Hi-C 172\,{\AA}.
To highlight fainter structures, the IRIS image is on a logarithmic scale, the Hi-C image is scaled with a power of 0.3. The multi-color composite of these two bands is displayed in panel (d), with green for Hi-C and red for IRIS data.
For context, in panel (c) we display a co-temporal magnetogram acquired by HMI showing the line-of-sight component of the magnetic field in the photosphere ($\pm$200\,G).
The footpoints of features [1] and [2] are indicated by the arrows. The rectangle highlights the peculiar small loop [3] that is briefly discussed in \app{S:short}.
The Hi-C image shown here was taken at 19:00:33 UT, the IRIS and HMI data are those available closest in time.
North is top and west is right. 
\label{F:hic.core}}
\end{figure*}

This active region was the target of the High-resolution Coronal imager (Hi-C 2.1) suborbital rocket experiment during its flight on May 29, 2018.
The details of the Hi-C experiment are described in \cite[][]{2019SoPh..294..174R}.
During our observation, the center of the active region was located at an
heliocentric angle of $\mu{=}\cos\vartheta$ of about 0.95, i.e., close to
disk center with an angle between the line-of-sight and the local vertical
of only about 15$^\circ$.
The data are freely available through the Virtual Solar Observatory (\url{https://www.virtualsolar.org}).

Between 18:56 and 19:02\,UT the Hi-C imager collected science data in a band centered around 172\,{\AA}.
The temperature response of this band is similar to 171\,{\AA\ channel of the Atmospheric Imaging Assembly, AIA \cite[][]{2012SoPh..275...17L}.
The bulk part of the plasma imaged by Hi-C is at temperatures just below 1\,MK.
The plate scale of the Hi-C data is 0.129{\arcsec} per pixel and the data have been taken with an exposure time of 2\,s and a cadence of 4.4\,s.
The spatial resolution in the sharpest frames from the Hi-C flight is estimated to be 0.35{\arcsec}.
The resolution varied over the flight due to blurring by jitter. 
In our study we concentrate on image frame \#\,58 taken at 19:00:33\,UT which is one of the sharpest images taken during the flight \cite[][]{2019SoPh..294..174R}.
Only when we analyze the temporal evolution in a single loop in \sect{sec:tevol} we utilize all Hi-C frames.

We relate the Hi-C observations to data from the Interface Region Imaging Spectrograph, IRIS \cite[][]{2014SoPh..289.2733D}, in particular those acquired in the 1400\,{\AA} band by the slit-jaw camera.
The bright features in this band in active regions represent plasma at just below 0.1\,MK \cite[][]{2018ApJ...854..174T}.
The 1400\,{\AA} images have been taken with an exposure time of 2\,s and a cadence of 13\,s.
The plate scale of the IRIS slit-jaw images originally is 0.17{\arcsec} per pixel, but the data taken during the Hi-C campaign are binned by 2$\times$2 pixels.
In the end, the spatial resolution of the IRIS and Hi-C data is roughly comparable, while it is significantly better than AIA (plate scale of 0.6{\arcsec} per pixel).
While the field-of-view of the IRIS slit-jaw images is smaller than that of Hi-C, the IRIS data fully capture the core of the active region that we investigate here (\fff{F:hic.core}a).
The IRIS data are available through \url{http://iris.lmsal.com}. The Virtual Solar Observatory
also offers the IRIS data taken during the Hi-C 2.1 flight.

The proper spatial alignment of the data from Hi-C and IRIS is important for this study.
Thus we provide details of our alignment procedure in \app{sec:align}. Using data from AIA we can achieve a spatial alignment between Hi-C and IRIS better than
0.5{\arcsec}, corresponding to about 350\,km at disk center on the Sun. The data from AIA are available through the   Joint Science Operations Center (JSOC) at \url{http://jsoc.stanford.edu}.

For the context of the magnetic field structure we use data from the Helioseismic and Magnetic Imager, HMI \cite[][]{2012SoPh..275..207S}.
In \fff{F:hic.core}c we show the line-of-sight magnetic field in the core of the active region aligned with the data from Hi-C and IRIS.
Like AIA, the HMI data are available at JSOC. For an overview of the temporal evolution of the active region during the days leading up to the Hi-C observation we utilized the JHelioviewer software \cite[][]{2017A&A...606A..10M}.

\section{Parallel cool and warm loops in active region cores\label{S:results}}

The main  observational goal of this study is to relate the loops in the active region core seen in the images of Hi-C to those in IRIS.
These loops are the elongated features %
seen in both the Hi-C and the IRIS data (panels a, b of \fff{F:hic.core}).
Comparing with the magnetic field at the solar surface (panel c), mostly, the two footpoints of each loop feature are rooted in patches of opposite magnetic polarities, two examples are highlighted by arrows.
Of course, in many cases the moderate spatial resolution of HMI is not sufficient to resolve the magnetic structure at the loop footpoints \cite[][]{2017ApJS..229....4C}.
Still, this comparison underlines that most of these elongated features seen in extreme UV images are indeed plasma loops outlining the magnetic field structure.

\subsection{Loops in Hi-C and IRIS images}

Very often the loops in the core of the active region seen in Hi-C and IRIS data do run in parallel.
For a better spatial comparison between the two data sets we plot a multi-color composite of the Hi-C 172\,{\AA} channel and the IRIS 1400\,{\AA} slit-jaw image in \fff{F:hic.core}d.
Here one can easily identify many green and red features (representing Hi-C and IRIS respectively) that run parallel to each other over a considerable part of their length.
We highlight two structures by pointing with arrows to their end or footpoints.
Feature [1] is a single loop.
Feature [2]  forks in the middle, so that it has two endpoints on the eastern side and only one footpoint on the western end, indicating that is a small bundle of loops whose footpoints diverge slightly.
However, the IRIS images (see \fff{F:hic.cuts}) show a slightly different structure, with the common right part of the fork being bend more southward.

Depending on the quality of the spatial alignment of the data sets, it is possible that the loops seen in Hi-C and IRIS either run in parallel with a small offset, or that they are at exactly the same location.
These two possibilities would imply two rather different scenarios.
In the former case, the loops we see in Hi-C and IRIS have to be different entities at different temperatures that are separated in space.
In the latter case, the Hi-C and IRIS loops might be actually the same structure that would have to consist either of one single temperature or be a multi-stranded multi-temperature loop, with the strands unresolved by either instrument.
If the loops appear at the same location there is also the possibility that they are crossing the same line of sight with some spatial offset along the line of sight.
This would apply in special situations only, because in general we cannot expect that multiple solar structures will be aligned with the line of sight along their entire length.
A careful spatial alignment is important for our study and we describe the alignment procedure and its test in \app{sec:align}.

\begin{figure*}
\centerline{\includegraphics[width=\widcuts]{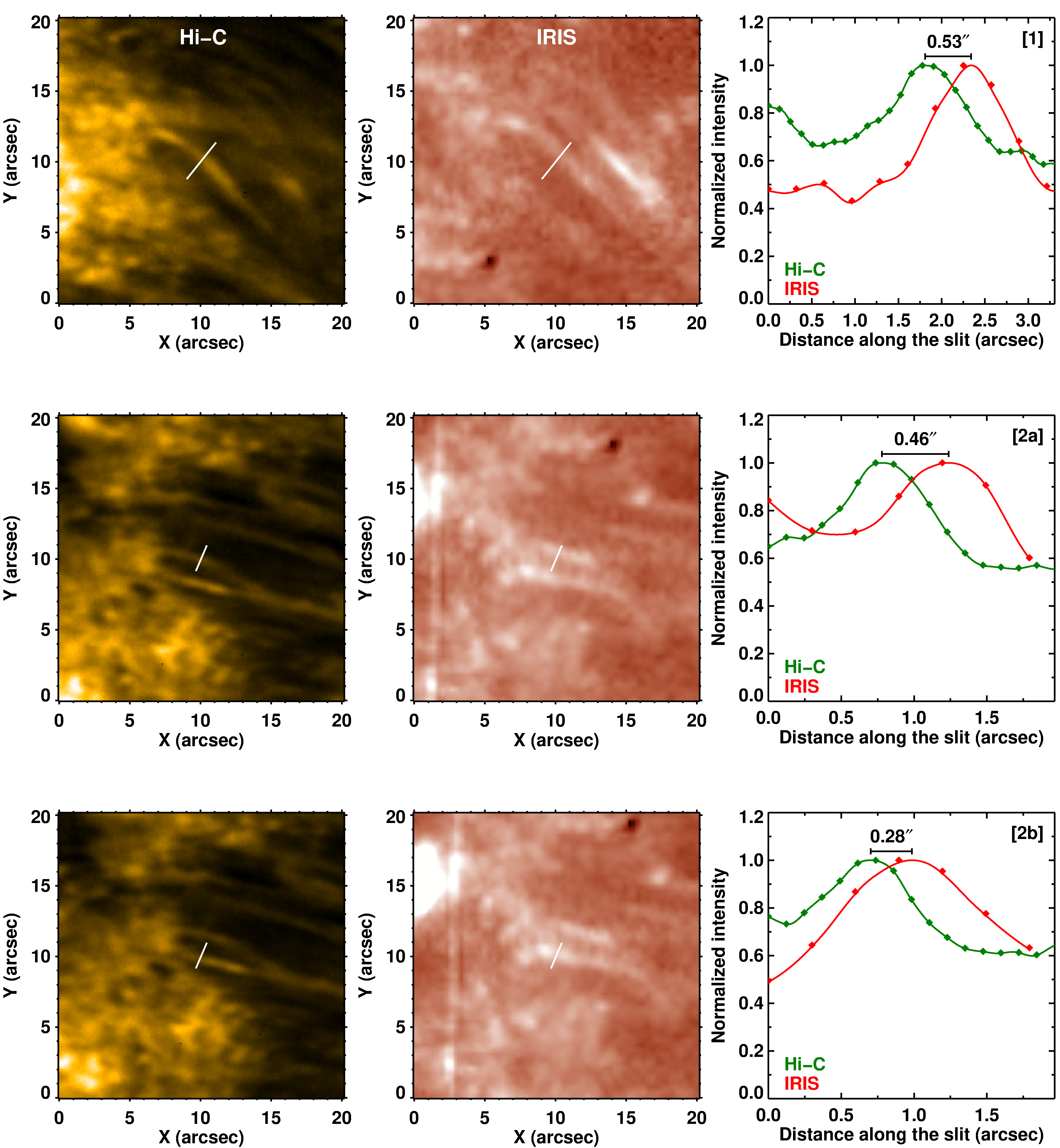}}
\caption{Spatial offset between loops seen in Hi-C and IRIS.
The left and middle panels show zooms into the Hi-C and IRIS images centered around the features as marked in \fff{F:hic.core}.
The top row shows the loop feature [1] the middle and bottom row the two parts of the double feature [2], here marked [2a] and [2b].
In the respective right panel we display the variation of the intensities in Hi-C and IRIS along the white lines (strips) marked in the respective left panels, i.e.\ roughly perpendicular to the loop.
The diamonds show the data interpolated on the plate scale of the  respective instrument, the solid curves show interpolations at higher resolution to measure the offsets between the two instruments.
The offsets between Hi-C and IRIS are marked with the plots.
The typical error for the offsets is 0.1{\arcsec}.
See \sect{S:offset.across}.
\label{F:hic.cuts}}
\end{figure*}

\subsection{Spatial offset between IRIS and Hi-C\label{S:offset.across}}

To investigate the offset between the loops seen in Hi-C and IRIS we plot cuts across the two selected loop features in \fff{F:hic.cuts}.
There we show the intensity variation along the strip across the loop interpolated onto the plate scale of Hi-C and IRIS. We also show intensities interpolated to a higher resolution, which is used to determine the position of the loops and their offset.
In the case of the forking feature [2] we select two locations where it appears as a double structure.
We determine the offset between the loops through the location of the peak of the respective cross-sectional intensity cuts (\fff{F:hic.cuts}).

Because the profiles are quite smooth, we can determine the location of the features and thus their offset with an accuracy that is a fraction of one spatial resolution element (typically the error is 0.1{\arcsec}, see \app{sec:samples}).
As noted with \fff{F:hic.cuts}, the offsets between the loops seen in Hi-C and IRIS are 0.53{\arcsec} for feature [1] and 0.28{\arcsec} and 0.46{\arcsec} for the double-feature [2].
Even if we were to change the absolute alignment between the Hi-C and the IRIS data, we would not be able to overlay all three cross-sections simultaneously.
Investigating more examples in \app{sec:samples} confirms this finding; we find displacements between Hi-C and IRIS
ranging from 0{\arcsec} to more than 0.7{\arcsec} 
(see \fff{fig:ex.loops}, \tab{T:offsets} and \fff{fig:offsets}). 
Most importantly, one cannot shift (or align) the IRIS and Hi-C data in a way that all the features would overlap. No matter how the images are shifted with respect to each other, there would be an offset of at least more that 0.5{\arcsec} for some of features in Hi-C and IRIS.
Thus we conclude that the offset between the loops seen in IRIS and Hi-C is real.
Typically, this small displacement between the loops is smaller than or comparable to their widths.

The loops seen in Hi-C and IRIS are not only at different locations, they are also composed of plasma at different temperatures. While the emission seen in the IRIS images is dominated by \ion{Si}{4} forming around 0.1\,MK, the analysis in \sect{S:res.temp} shows that the emission seen by Hi-C in these loops originates from temperatures around 0.3\,MK (and in some cases maybe up to almost 1\,MK).
The temporal evolution of the loops seen in IRIS and Hi-C is mostly (but not always) different (\sect{S:res.time}).

\subsection{Spatial offsets along the loops\label{S:offset.along}}

The multi-color image in \fff{F:hic.core} gives a first indication that the loops run in parallel.
To quantify this more, we show in \fff{F:hic.along.loop} how the offset between a loop seen in Hi-C and IRIS changes along the loop.
Here feature [1], which is clearly discernible in both the Hi-C and IRIS images, shows an offset that is mostly constant over the better part of the length of the loop.
Thus here the loops in Hi-C and IRIS are parallel.

Of the eleven cases listed in \tab{T:offsets} and discussed in \sect{S:offset.across} and \app{sec:samples}, eight show a variation similar to feature [1], some as clear, some a little less clear.
Only three cases show a variation of the offset, with changing signs, along the loop, meaning that they do not clearly show the offset towards the same side (features 2a, 10, 11).

An intriguing case is the combination of features 2a, 2b and 8.
Together they seem to form a forking structure (bottom two panels of \fff{F:hic.cuts} and \fff{fig:ex.loops}[8]).
One could now speculate that this is a braided structure similar to the magnetic field lines in \fff{F:braid.fork}, in particular because the offsets of the left side of the fork for [2a] and [2b] are in the opposite direction than for the right side with feature [8] (see also \fff{fig:offsets}).
However, on the one hand it is not clear if this forking structure is one single feature or the line-of-sight composition of different loops.
On the other hand, an apparent forking feature is not necessarily braided, as can be nicely seen in the numerical models of \cite{2021arXiv210410940C} with their Figs.\ 4b, 4d and 5.

\begin{figure}
\centerline{\includegraphics[width=75mm]{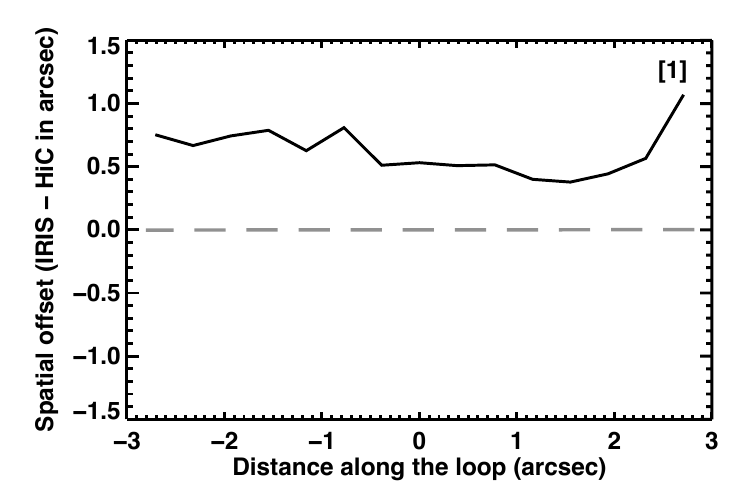}}
\caption{Offset between loop in Hi-C and IRIS data.
This shows the offset along feature [1] pointed to in \fff{F:hic.core} and shown in \fff{F:hic.cuts}.
The distance 0{\arcsec} along the loop is at the location in \fff{F:hic.cuts}[1] where the white slit crosses the loop.
The offset between Hi-C and IRIS is consistently to the same side.
See \sect{S:offset.along}.
\label{F:hic.along.loop}}
\end{figure}

\section{Temporal evolution and thermal structure of the loops\label{sec:tevol}}

From \sect{S:results} and \app{sec:samples} we conclude that  the loops seen in Hi-C and in IRIS are not at the same spatial location. Here we investigate the evolution of these loops in space and time (\ref{S:res.time}) as well as their temperatures (\ref{S:res.temp}). If these properties are different for the loops in Hi-C and IRIS, it would be further support that features seen in the two instruments are different.

\begin{figure}
\centerline{\includegraphics[width=\widex]{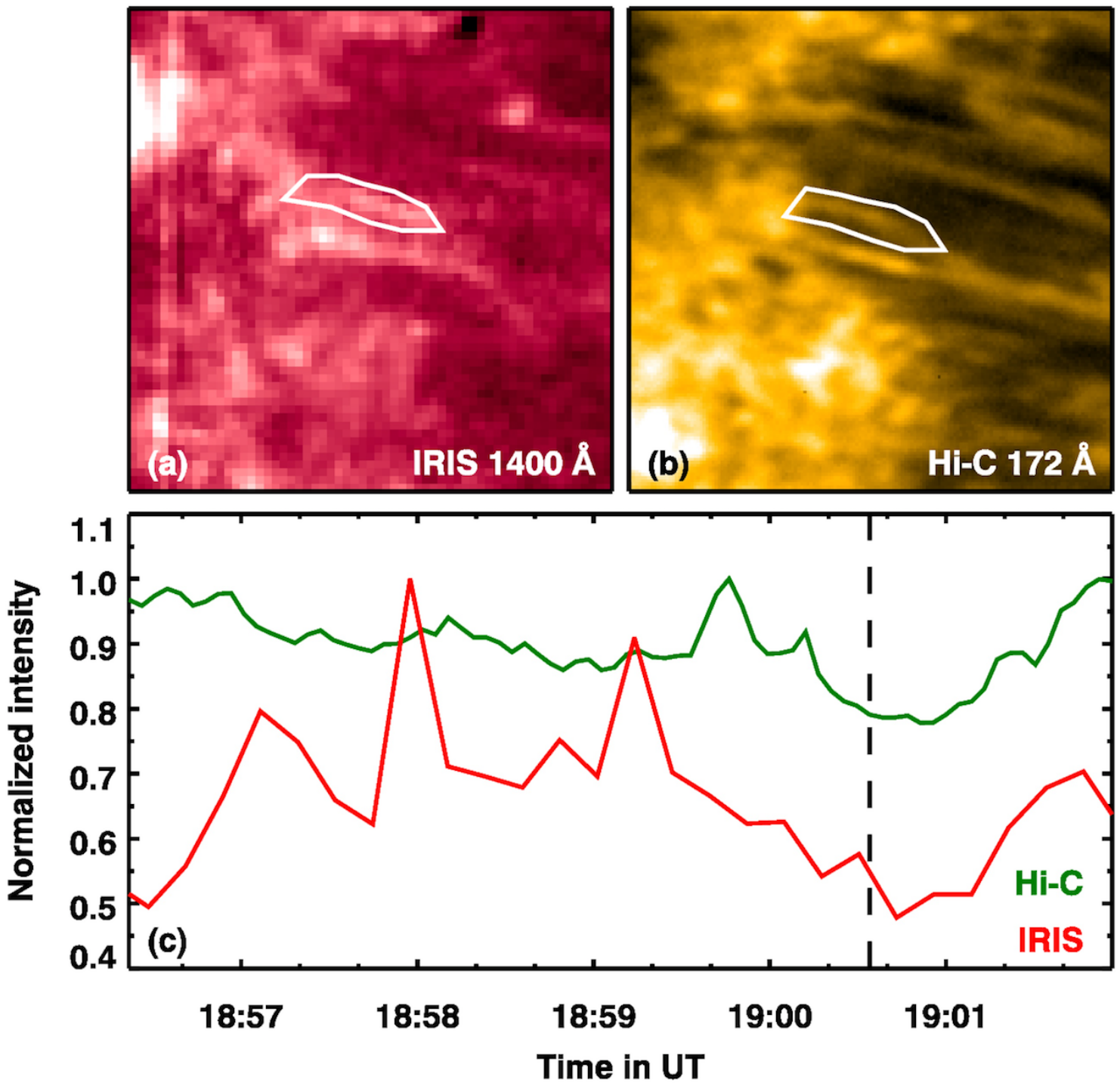}}
\caption{Spatio-temporal evolution of the intensity in one selected loop feature.
Panels (a) and (b) show a zoom of the IRIS 1400\,{\AA} and Hi-C 172\,{\AA}
channels into feature [2a] identified in \fff{F:hic.core}.
Panel (c) displays the temporal evolution in part of the respective feature
integrated over the polygons indicated in the top panels.
These polygons encompass 74 and 514 pixels in the cases of IRIS and Hi-C,
respectively.
The light curves in Hi-C (green) and IRIS (red) are shown in normalized to their peak value.
The peak counts are about 3600\,DN for IRIS and 3.4$\times$10$^6$\,DN in
Hi-C.
The dashed line in panel (c) indicates the time the images in panels (a)
and (b) are recorded.
A movie showing the temporal evolution is available online (\url{http://www2.mps.mpg.de/data/outgoing/peter/papers/2019-hic/movie-s1.mp4}).
See \sect{S:res.time}.
\label{F:hic.time}}
\end{figure}

\subsection{Spatio-temporal evolution of Hi-C and IRIS loops\label{S:res.time}}

To analyze the evolution of the loops we follow the features as seen in Hi-C and in IRIS in time. Because Hi-C is a sub-orbital rocket experiment, this analysis is limited to about five minutes.
Here we first concentrate on feature [2] marked in \fff{F:hic.core}. In \fff{F:hic.time} we show stills from the Hi-C and IRIS data along with the light curves of the northern fork of this feature (the movie attached to \fff{F:hic.time} illustrates the temporal evolution). To enhance the signal-to-noise level of the temporal evolution and in order not to depend on variations on small (pixel) scales, we integrate the intensity in each band over a good part of this upper fork of the loop, marked by the polygons in \fff{F:hic.time}a,b.
We do not subtract a background, because the region is quite complex and the background subtraction would introduce artifacts.
The relative variability we see in both Hi-C (up to about 10\%) and in IRIS (up to 30\%) is significant.
Considering the count rates (in the spatially integrated signal; see \fff{F:hic.time}) we estimate that the uncertainties of the intensities should be one percent or less for both Hi-C and IRIS.

The temporal variability in the Hi-C and IRIS data is different.
To begin with, the relative variations are much larger in the IRIS data.
Based on cooling time scales, one would expect a higher temporal variability for emission lines forming at lower temperatures (towards the transition region) as compared to those forming in hotter regions.
This is because radiative and conductive cooling times get longer with increasing temperature \cite[e.g.,][Sect.\,4.3.1]{Aschwanden.2005}.
Therefore the higher variability in IRIS suggests that the IRIS loops are somewhat cooler than the loops seen in Hi-C.

In addition to differences in the overall evolution, short-term brightenings lasting only a fraction of a minute are not related to each other in the two bands.
This is also supported by the visual impression when investigating the movies of the IRIS and Hi-C data (see movie attached to \fff{F:hic.time}).
In general, in the literature there are reports of cases where (roughly) co-spatial loops seen in the IRIS 1400\,{\AA} band and in the AIA 171\,{\AA} band (similar to Hi-C) are correlated in their temporal evolution.
In some examples time delays between the two bands have been observed suggesting heating of the plasma in one single (multi-stranded) structure \cite[][]{2019A&A...626A..98L}.
The case we show here in this study is different.
In our case, the time evolution in Hi-C and IRIS seems to be disconnected.
Together with the finding of the spatial offsets, this suggests that the loops we see in Hi-C and IRIS are definitively disconnected, even though they might be hosted in the same bundle of emerging field lines.

In some cases there are also similarities between the light curves between IRIS and Hi-C. In feature [2a] in \fff{F:hic.time} the intensity in both instruments increases during the last minute.
However, during the first 4 minutes, they differ significantly.
Firstly, as mentioned above, the short brightenings in IRIS (at 18:58 and just after 18:59) have no counterparts in Hi-C.
Secondly, when discarding these short peaks, during the first 3.5 minutes the IRIS curve would increase by more than 50\% (in normalized intensity from 0.5 to 0.75) before it drops again (to 0.3 at 19:00:30).
During this time the variability in Hi-C is only 10\% at a rather constant level.
Looking at the movie attached to \fff{F:hic.time} it is also clear that the spatio-temporal evolution of the two loops in Hi-C and IRIS is quite different. Blobs appear in both the IRIS and Hi-C loops, but at different locations along the loop and at different times.
Therefore, this provides support to the idea that the temporal evolution seen in this feature [2a] is quite different in Hi-C and IRIS, despite some similarities.

This case of feature [2a] is representative: mostly, the loops show a different evolution in Hi-C and IRIS, in both time and space.
In \fff{fig:ex.tevol} we show the temporal evolution for more examples, now with a broader coverage in temperature by using data from AIA.
Because the variation in Hi-C is very similar to the 171\,\AA\ channel of AIA (and to keep the figure more clean) we do not add Hi-C in \fff{fig:ex.tevol}. In general, the loops in the active region core brighten over time scales on the order of a fraction of a minute to a few minutes.
Following the evolution in AIA for longer than the duration of the Hi-C flight, the loops seem to be present at the same location for longer times. This could be either because the same structure brightens up again, or that a different strand appears.
The loops we study here are probably related to inter-moss loops \cite[][]{2013ApJ...771...21W}: at least one feature shows a very similar evolution in the AIA channels as the inter-moss loops (cf.\ \fff{fig:ex.tevol} feature [5]), the other cases are less clear. Furthermore, those inter-moss loops have com\-parable length and lifetime as the loops studied here.

Based on the comparison of the spatio-temporal evolution of the loops seen in Hi-C and IRIS we conclude that they are sometimes but not always identical. Still, the question remains if they might have the same or different temperatures which we address in the following \sect{S:res.temp}.

\begin{figure*}
\includegraphics[width=\textwidth]{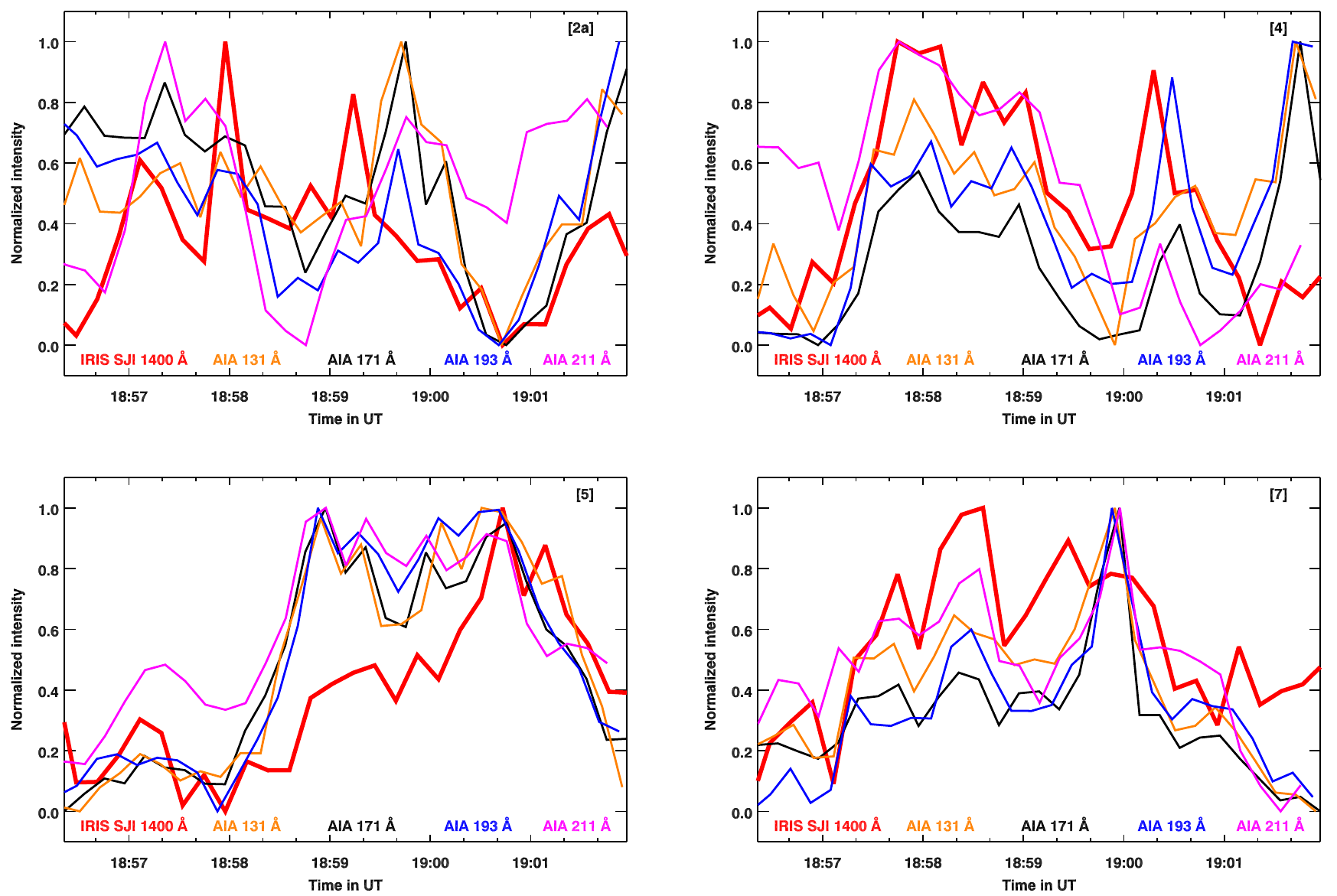}
\caption{Temporal evolution of selected loops.
Each panel shows the light curves for one loop example for four AIA channels
and the IRIS 1400\,{\AA} images.
The number of each feature is given in the top right corner of each panel.
The temperature contributions (or kernels) of these AIA channels and the
\ion{Si}{4} line are displayed in \fff{fig:contrib.fct}.
The labels in the panels refer to the features shown in \figs{F:hic.cuts}
and \ref{fig:ex.loops}.
In order not to overcrowd the panels, we do not show the light curves in
the Hi-C 172\,{\AA} band which are almost exactly the same as for the AIA
171\,{\AA} channel.
All light curves are normalized by mapping the respective minimum and maximum
values to the range of [0,1].
See \sect{S:res.time} and \app{S:res.temp.detail}.
\label{fig:ex.tevol}}
\end{figure*}

\subsection{Temperature of the loops seen in Hi-C and IRIS\label{S:res.temp}}

Here we investigate the temperatures the loops in the active region core seen by Hi-C.
The availability of AIA data with a broader temperature coverage can provide some information on the temperature of the plasma from where the emission captured by Hi-C originates.
In particular we want to investigate if indeed the source region of Hi-C is at a higher temperature than the plasma seen with IRIS, i.e. that IRIS shows cool loops and Hi-C warm loops.

\subsubsection{Temperatures seen in extreme-UV channels}\label{S:temp.IRIS.AIA}

The emission recorded by the IRIS slit-jaw images in the 1400\,\AA\ channel is dominated by the \ion{Si}{4} doublet at 1394\,\AA\\ and 1403\,\AA, at least in the bright active-region features we investigate here \cite[cf.][]{2014SoPh..289.2733D,2018ApJ...854..174T}.
Under ionisation-equilibrium conditions \ion{Si}{4} originates from plasma just below 0.1\,MK, as illustrated by its contribution function shown in \fff{fig:contrib.fct} (red line).

\begin{figure}
\centerline{\includegraphics[width=\widex]{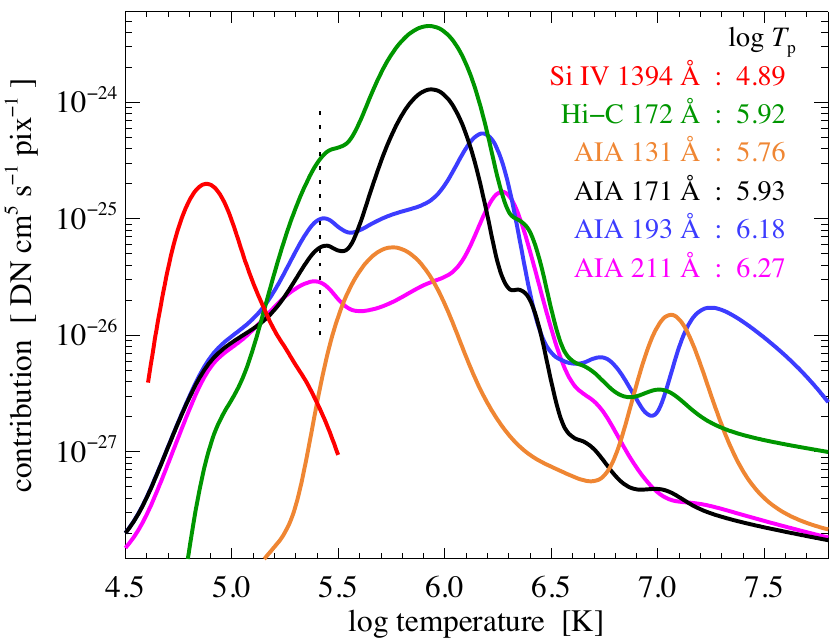}}
\caption{Contribution functions for Hi-C, AIA and \ion{Si}{4}.
The response functions for the (selected) AIA channels are calculated using the routines available in SolarSoft (\url{http://www.lmsal.com/solarsoft/}). 
The corresponding curve for Hi-C is from the Hi-C instrument paper \cite[][]{2019SoPh..294..174R}.
The contribution function for the \ion{Si}{4} line is calculated using Chianti \cite[][]{2019ApJS..241...22D} and is scaled to fit on this plot.
For each of the curves the temperature $T_{\rm{p}}$ of the peak of the contribution function is listed with the legend as $\log_{10}\,T_{\rm{p}}$\,[K].
The dotted line indicates a temperature of 0.26\,MK.
See \sect{S:temp.IRIS.AIA} and \app{S:temp.evol.contrib}.
\label{fig:contrib.fct}}
\end{figure}

For the 172\,\AA\ channel of Hi-C the situation is a bit more complex.
Besides the main peak just below 1\,MK, the Hi-C contribution function shows also a significant bump at lower temperatures at about 0.3\,MK (as indicated by the dashed line in \fff{fig:contrib.fct}).
So, depending on the density and temperature distribution along the line of sight, the emission seen in Hi-C could originate from plasma at temperatures somewhere between 0.3\,MK and 1\,MK. The situation is similar for many of the extreme UV channels of AIA, where the channels at 171\,\AA, 193\,\AA, and 211\,\AA\ show even a side peak at about 0.3\,MK (see \fff{fig:contrib.fct}).

Comparing the temporal variability in the different EUV channels, \cite{2013ApJ...771...21W} agued that one can distinguish whether a particular brightening is originating from about 0.3\,MK or just below 1\,MK. We apply their arguments to the small loop structures investigated here and describe this in \app{S:temp.diag}.
In the end, it does not really matter if the source region of Hi-C is mostly at around 0.3\, MK as would be argued following \cite{2013ApJ...771...21W}, or closer to 0.8 MK as one would judge from the the peak of the contribution function alone (cf.\ \fff{fig:contrib.fct}).
The main conclusion is that in some cases the source region of Hi-C is hotter than the plasma IRIS sees, so that IRIS shows cool loops while Hi-C shows warm loops.

\subsubsection{Dichotomy of temperatures in loops seen by Hi-C and IRIS}

The above result indicates that in some cases there is some observational dichotomy for the presence of plasma around 0.1\,MK (seen in IRIS 1400\,\AA) and at higher temperatures (seen in Hi-C\,172 and the AIA channels).
There will be also plasma at temperatures in-between, as seen in the 3D MHD model (\sect{S:model}), but plasma at those temperatures is not contributing (significantly) to the observables that we have available for analysis here, i.e. the images in the IRIS 1400\,{\AA} and the Hi-C 172\,{\AA} bands.

\subsection{Summary of spatial, temporal and thermal structure\label{S:res.summary}}

Combining the results on the spatial offsets of the loops seen in Hi-C and IRIS (\app{sec:samples}) with those on their spatio-temporal evolution (\sect{S:res.time}) and temperature (\sect{S:res.temp}) we conclude that at least some of} the loops seen in Hi-C and IRIS are different entities.
They appear at slightly different location, with an offset being smaller than their width, have different evolutions, and are at different temperatures.

This suggests that the loops seen in Hi-C and IRIS are neighboring strands at (slightly) different temperature in a bundle of magnetic field lines.
The field lines in this bundle cannot be twisted significantly on observable scales because the loops in Hi-C and IRIS are essentially parallel.
Some of the loops in Hi-C and IRIS loops overlap partly, i.e.\ their distance is smaller than their width, which is indicative of a gradual change of temperature perpendicular to the magnetic field, so that the source region of the lines contributing to the different instrument bands are overlapping in space.
This implies that there should be no (significant) twist of the magnetic field even within each loop. 
These conjectures are supported by the model presented in the next section.

\begin{figure*}
\centerline{\includegraphics[width=120mm]{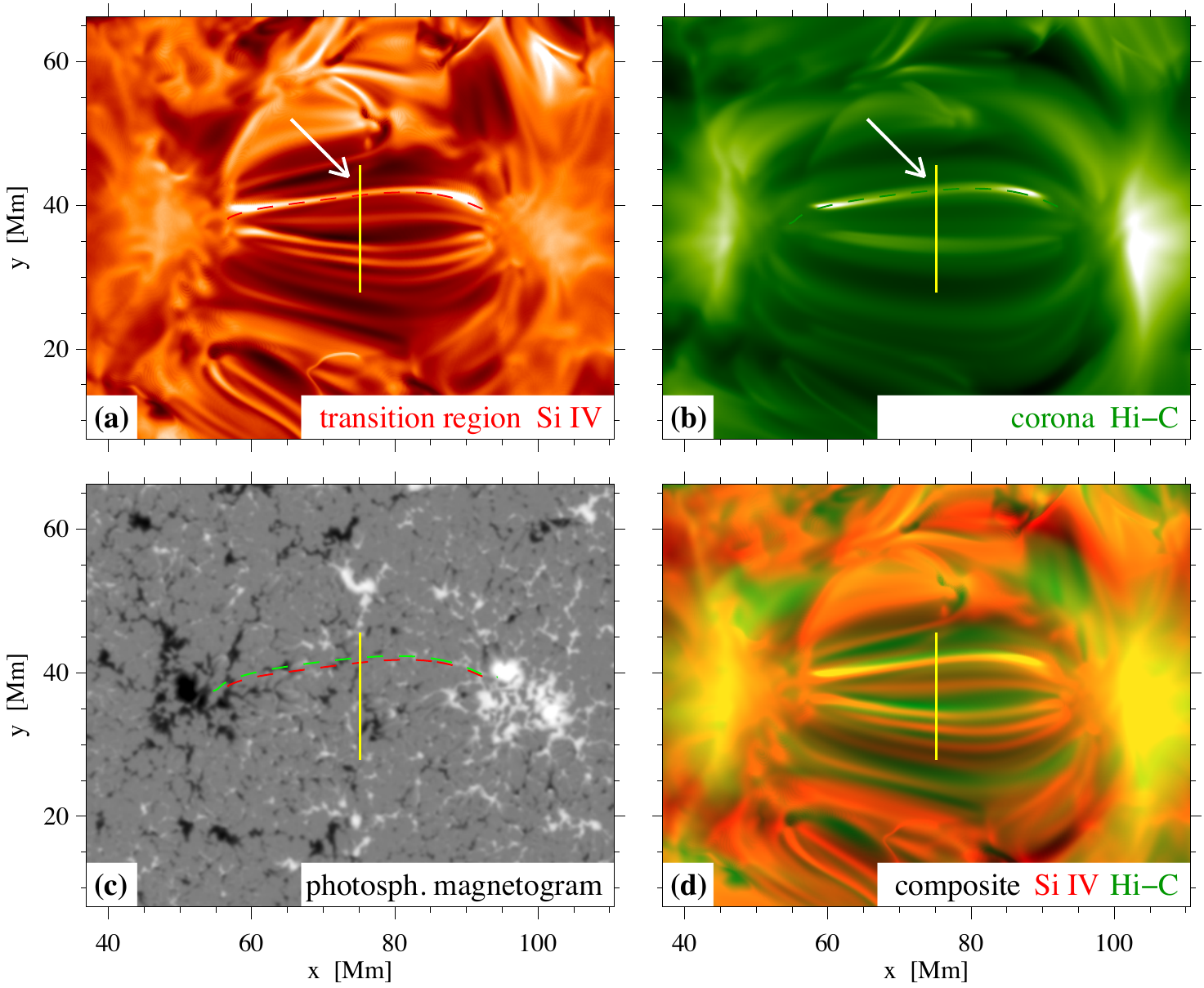}}
\caption{View of the 3D MHD model from straight above.
The layout is similar to the observations in \fff{F:hic.core}.
Panels (a) and (b) show the emission originating from \ion{Si}{4}, similar to the IRIS 1400\,{\AA} image, and from Hi-C.
Both the \ion{Si}{4} and Hi-C intensity have been integrated vertically through the computational domain.
Panel (c) shows the vertical component of the magnetic field in the photosphere of the simulation.
The multi-color composite of the cool and warm emission is displayed in panel (d).
Here the red channel represents \ion{Si}{4}, the green channel Hi-C.
The field of view covers only about one quarter of the horizontal extent of the 3D MHD model.
The red and green dashed lines show the projection of magnetic field lines passing through the center of the loops seen in \ion{Si}{4} and Hi-C (traced in the 3D data cube).
The vertical yellow lines shows the location of the cuts in intensity shown in \fff{F:model.cut} and of the vertical cuts in \fff{F:model.vert.cut}.
The arrows in panels (a) and (b) point to the respective loop in \ion{Si}{4} and Hi-C highlighted also by arrows in \figs{F:model.cut} and \ref{F:model.vert.cut}.
See \sect{S:res.model}. 
\label{F:model.img}}
\end{figure*}

\section{3D MHD model of loops in an emerging active region\label{S:model}}

To improve the interpretation of our observations we use a three-dimensional magneto-hydrodynamic (3D MHD) coronal model of an emerging active region.
In that model a new active region forms in response to the large-scale emergence of magnetic flux resulting in two sunspots and a complex evolving magnetic field between and around the spots.
During the flux emergence process the plasma along the rising magnetic field lines is heated and forms a system of coronal loops \cite[][]{2014A&A...564A..12C}.
The loops are energized by surface motions: the footpoints of magnetic field lines are dragged by magneto-convective flows, often toward the sunspots
 \cite[][]{2015NatPh..11..492C}.
This is similar to the flux-tube tectonics scenario \cite[][]{2002ApJ...576..533P}.
This model is thought mainly to illustrate a possible magnetic configuration of how the loops at different temperatures relate to each other spatially.
As such, the model output should be considered as a guide for the interpretation, while the specific spatial or temporal variation might not necessarily fit exactly the observations.

\subsection{Basics of 3D MHD model\label{S:model.basics}}

The 3D MHD model solves the mass, momentum, and energy balance along with the induction equation \cite[][]{2011A&A...530A.112B}, and uses a super-time-stepping scheme to improve the computational speed.
The energy dissipated through Ohmic and viscous dissipation is redistributed through field-aligned heat conduction and is lost through optically thin radiative losses.

For our study we re-analyze a high-resolution version of the aforementioned 3D MHD model \cite[][]{2015A&A...581A.137C}.
This covers the emerging active region with a computational domain spanning over about 147\,Mm $\times$ 74\,Mm horizontally.
The grid spacing (in the horizontal directions) is about 145\,km, which corresponds to about 0.2{\arcsec} at disk center for Earth-based observations.
This is of the order of the spatial resolution of the observations we use here (cf.\ \sect{S:obs}).

Using the temperature and density provided by the MHD model, we synthesize optically thin emission lines in the 3D data cube assuming ionization balance and electron collisional excitation \cite[][]{2004ApJ...617L..85P,2006ApJ...638.1086P}.
To calculate the extreme UV images we integrate the synthesized intensity along the vertical direction.
This corresponds to an observation of an active region from straight above, i.e., near disk center (as is the case for the active region we observe).
In this model part of our study we use the emission line of \ion{Si}{4} which is dominating the IRIS 1400\,{\AA} channel when observing bright transition region structures in an active region, as it is the case here.
To synthesise the emission as seen in the Hi-C 172\,{\AA} band we employ the Hi-C temperature response function as published in the instrument paper \cite[][]{2019SoPh..294..174R}.
Because we do not perform a one-to-one quantitative comparison between observations and numerical model, we do not apply a point-spread-function or other instrumental effects to the synthesized data.

\subsection{Cool and warm loops in 3D MHD model\label{S:res.model}}

The 3D MHD model produces an active region with numerous loops over a range of temperatures in its core, i.e.\ between the sunspots.
We show the core of that simulated active region in \fff{F:model.img} as it would appear when observed from straight above.
With 73\,Mm $\times$ 59\,Mm this represents about one quarter of the horizontal extent of the computational domain.
These simulation data are shown in the same format as the observations in \fff{F:hic.core}.
The loops in \ion{Si}{4} (similar to IRIS 1400\,{\AA}) and Hi-C are displayed in panels (a) and (b) of \fff{F:model.img}.
A composite image of these two in panel (d) shows how they relate to each other.

The loops in the core of the active region in this simulation are longer than those in the observations.
The modeled loops have lengths of some 30\,Mm to 40\,Mm, while in the loops seen in the Hi-C observations have lengths of up to 15{\arcsec}, corresponding to about 10\,Mm.
Still, model and observations are comparable in that in both cases magnetic field is emerging giving rise to the formation of these loops.
In the model the scale of the flux emergence is larger, but we can expect the processes to be similar if the flux emergence occurred on a smaller scale, comparable with the observations.
Shorter loops (at lower heights and thus higher densities) would require substantially more heating to balance the radiative losses (that increase with density squared). MHD models for different setups produced loops with lengths of 20{\arcsec} reaching temperatures of well in excess of 1\,MK \cite[][]{2013A&A...555A.123B}.  Thus it should be possible that emerging loops with 15{\arcsec} length as observed here might reach temperatures in the MK range.   
Therefore we consider the difference between 3D model and observations to be quantitative but not qualitative.

\begin{figure}
\centerline{\includegraphics[width=80mm]{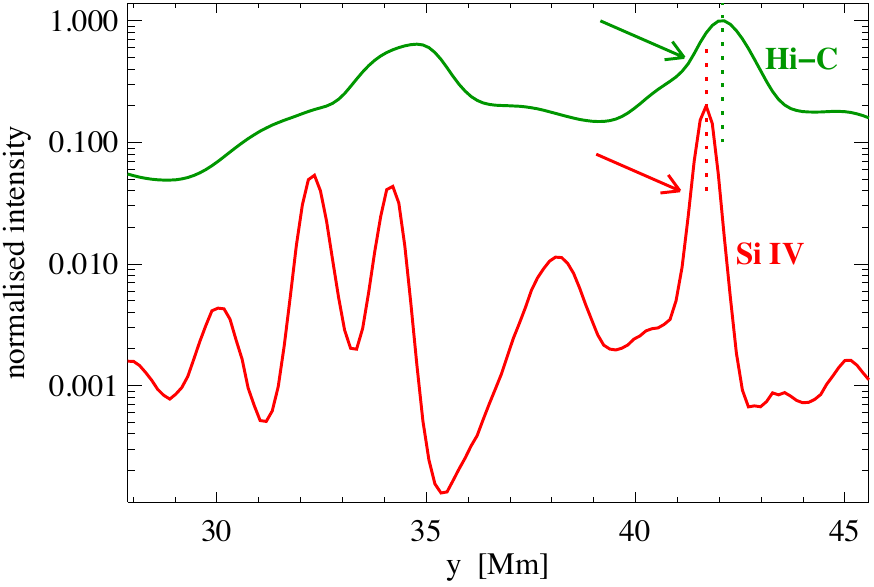}}
\caption{Cut through cool and warm loops in the 3D MHD model.
This shows the intensity variation in \ion{Si}{4} and Hi-C along the yellow line in \fff{F:model.img}.
The offset between the loops in \ion{Si}{4} and Hi-C near $y{\approx}42$\,Mm is about 400 km.
These loops are highlighted by arrows and are the same as pointed to in \figs{F:model.img} and \ref{F:model.vert.cut}.
See \sect{S:res.model}.
\label{F:model.cut}}
\end{figure}

As in the observations, the \ion{Si}{4} and Hi-C loops run in parallel.
One particular example for parallel cool and warm loops is the structure at around $y{\approx}42$\,Mm marked by arrows in \fff{F:model.img}.
The cut across these loops shown in \fff{F:model.cut} highlights that the 3D MHD model shows a spatial offset between loops in  \ion{Si}{4} and Hi-C,  just like in the observations.
In this example the loop separation of about 400\,km roughly corresponds to about 0.5\arcsec.
In general, the separation of the loops at different temperature will depend on the viewing angle (cf.\ \fff{F:model.vert.cut}c), and the good quantitative match found here is largely by chance.

In contrast to the observations, in the 3D MHD model we can investigate what causes this offset.
For this we first look at a 3D view of the simulation box in \fff{F:model.vert.cut}c.
Clearly, the loops in \ion{Si}{4} and Hi-C show a significant spatial offset, in both the horizontal and the vertical directions.
Checking the vertical cut through the the midplane of the loop in \fff{F:model.vert.cut}a reveals that the loops in \ion{Si}{4} and Hi-C are offset in the vertical direction by about 2\,Mm.
More importantly, they are also at different temperatures.
As expected, the \ion{Si}{4} loop appears at about 0.1\,MK and the Hi-C loop at just below 1\,MK.
So in the 3D MHD model we have a clear distinction between a cool (order 0.1\,MK) loop and a warm (order 1\,MK) loop.

\begin{figure*}
\centerline{\includegraphics[width=\textwidth]{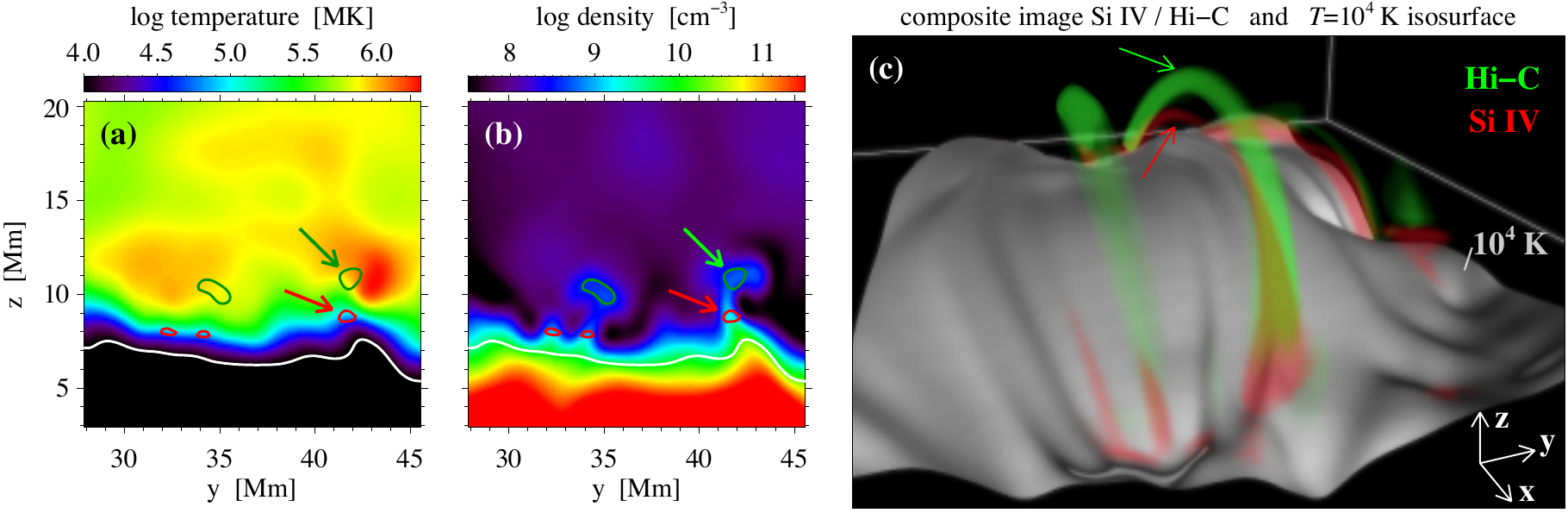}}
\caption{Structure of the loops in the 3D MHD model.
A vertical cut through the midplane of the loop(s) is shown in panels (a) and (b) for the temperature and the density.
The vertical cut is located at the yellow line in \fff{F:model.img}
The white line indicates the isocontour of a temperature of $10^4$\,K. The red and green lines show isocontours of the emission in \ion{Si}{4} and Hi-C.
Essentially they show the location where the cool and warm loops cross the midplane.
Panel (c) shows a 3D visualization of the cool and warm loops seen in \ion{Si}{4} and Hi-C together with a temperature isosurface at $10^4$\,K. The arrows point at the same cool and warm loops as in \figs{F:model.img} and \ref{F:model.cut}.
See \sect{S:res.model}.  
\label{F:model.vert.cut}}
\end{figure*}

There is a significant difference between the warm and the cool loop, though.
The warm loop is a structure in a high-temperature and high-density region that is a distinct feature in the upper atmosphere (cf.\ \fff{F:model.vert.cut}a,b).
The density at the location of the warm loop synthesised for Hi-C is higher than in the regions directly below, which can be seen clearly in particular for the Hi-C loop at $y{=}35$\,Mm and $z{=}10$\,Mm in \fff{F:model.vert.cut}b. 
This is what one might call a proper coronal loop.
In contrast, the cool transition region loop sits just above the chromosphere which is itself elevated by the emerging magnetic flux.
In both the vertical cuts (\fff{F:model.vert.cut}a,b) and in the 3D view (\fff{F:model.vert.cut}c) the $T{=}10^4$\,K isocontour, indicating the top of the chromosphere, is clearly elevated under the coronal loop which is because the emerging and rising field lines push up the chromosphere. 
In some sense the cool transition region loop could be considered as grazing the top of the chromosphere.

In this 3D MHD model the magnetic field expands into the upper atmosphere, and as it expands, the plasma trapped on the field lines gets heated \cite[][]{2015NatPh..11..492C}.
The cool and the warm loops represent two (transient) stages of the evolution of an emerging bundle of magnetic field lines.
Still, at each given time the cool and the warm loops are aligned with the magnetic field, i.e., there is a field line that runs (roughly) along the center of the respective plasma loops.
Most importantly, this shows that (at least in this model), the warm and cool loops are different structures originating from different spatial locations.
To some extent, this distinction between cool and warm loops is an artifact of our limited observational capabilities.
If we had imaging instruments with a fine temperature coverage between 0.1\,MK and 1\,MK, essentially we would see a continuum of loops with increasing temperature side-by-side.
Using synthesized emission for a sequence of ions from \ion{Si}{4} through \ion{C}{4}, \ion{O}{4}, \ion{O}{5}, \ion{O}{6}, \ion{Ne}{8}, and \ion{Mg}{10} we indeed find such an overlapping continuous distribution of loops.

The cool and warm loops are part of the same emerging bundle of field lines, but they form in different places of this bundle, clearly separated spatially.
In the model this difference originates from the distribution of upward-directed Poynting flux at the surface.
Like in the flux-tube tectonics or the braiding scenario, the magneto-convection moves around patches of magnetic flux.
During the emergence process this can temporarily create small regions that have an enhanced upward Poynting flux at a fixed location, like a hot spot \cite[][]{2014A&A...564A..12C,2015NatPh..11..492C}.
This structure in the Poynting flux at the solar surface then results in the loops being heated while expanding into the upper atmosphere, so that at each given time the low-lying field lines would host cooler plasma while those field lines already emerged to larger heights carry hotter plasma.
Hence we see the cool loops just above the chromosphere and the warm loops at larger heights.
Because of the complex magnetic structure, the loops are free to move in all three directions as they expand, resulting in vertical and horizontal offsets.
If the emerging bundle of field lines would be strongly twisted and braided, then the field lines hosting the cool and warm loops, respectively, would be wound around each other. Not seeing that the cool and warm loops are wound around each other implies that the magnetic field relaxes faster than the field is stressed. This prevents braiding on length scales that are comparable to the width of the bundle of emerging field lines, viz. the spatial separation of the cool and warm loops.

In conclusion, the 3D MHD model of an emerging active region shows similar loop features as we find in the observations.
The magnetic structure that is modeled here is of case (1) as defined in \sect{S:braid}, i.e.\ a bundle of field lines that is rooted at each end in one single region.
Plasma at  different temperatures is seen in loops that run in parallel with a small offset. In the model the (large) dissipation leads to a relaxation that is faster than the braiding.
Consequently the emerging bundle of field lines does not show signatures of a highly braided state of the magnetic field.
Instead, the loops loaded with plasma at (slightly) different temperatures run in parallel.

As mentioned at the onset of this \sect{S:model}, this model is mainly thought to be an illustration of one possibility of how to have loops at different temperatures running in parallel with some offset (depending on the viewing angle).
Certainly it should not be considered as unique.
For example, the magnetic setup of the emerging low-lying loops  might not be typical for long stable loops that have been claimed  to be heated by braiding.
Also, if the emergence would be fast enough, braiding might not have sufficient time to create braided structures that could be visible in the observations.
Furthermore, the small-scale loops found in the center of the active region in the Hi-C and IRIS observations discussed in this study might have a different magnetic configuration.
The observed emerging loops might be energised by large-angle reconnection events that are triggered by the interaction between the emerging magnetic flux and the pre-existing coronal magnetic field \cite[e.g.][]{2016ApJ...825...93O}.
Despite of all these shortcomings, the model illustrates one way of creating parallel loops in IRIS and Hi-C observations.

\section{Discussion and conclusions\label{S:discussion}}

The key observational result of our study is that in most cases distinct loops (at different temperatures) are offset from each other and roughly parallel. 
Being part of the same bundle of (emerging) field lines, these show no sign of braiding or twisting. The spatial separation (perpendicular to line of sight) of the distinct loops is less than their width (cf.\ \fff{F:hic.cuts}). 
Therefore the lack of twisting between the loops  as seen in EUV emission implies that there is also no (noticeable) braiding of the field lines within each loop. 
Still, field-line braiding might be the source to energise the corona, but we simply do not see the observational signatures of this in the EUV imaging observations.

If the energisation of the corona is due to braiding of magnetic field lines, the observation of these parallel loops and the absence of braiding signatures has a significant implication: the magnetic field has to relax already while driven by horizontal motions at surface, and the field will not
reach a highly braided state \cite[][]{2007ApJ...662L.119S}.
This is consistent with current 3D models for the braiding of field lines: the coronal emission from a loop model driven by footpoint shuffling does not shows signatures of braiding \cite[][]{2016ApJ...817...47D}.
When self-consistently driven from the footpoints, the magnetic field relaxes before reaching a highly braided state  \cite[][]{2018A&A...615A..84R}.
In 3D MHD models of the corona \cite[][]{2010ApJ...718.1070H,2011A&A...530A.112B,2017ApJ...834...10R,2019A&A...624L..12W} computational constraints restrict the resolution so that the effective dissipation is rather large. This translates to (magnetic) Reynolds numbers in the models that are orders of magnitude smaller than expected for the solar coronal plasma \cite[][]{2015RSPTA.37350055P}.   While often considered as a shortcoming of the models, our results might hint that the (comparably) large dissipation in the upper solar atmosphere might be more realistic than generally thought, e.g.\ because  turbulence increases the effective dissipation \cite[][]{1984PPCF...26..311B,1996Ap&SS.242..165B}. Even under the presence of magnetic turbulence, the guide field is still rather smooth in the direction along the loop because the relaxation is as fast as (or even faster than) the driving \cite[][]{2018A&A...615A..84R}.
In the (upper) chromosphere ambipolar diffusion is very strong and plays a key role \cite[e.g.][]{2017Sci...356.1269M}. Hence there the magnetic Reynolds number can be represented by the numerical models.

Our observations of parallel loops in Hi-C and IRIS put a new constraint on  models for field-line braiding.
Despite the braiding and twisting of the magnetic field, the loops observable in EUV should not show twisted structures.
This condition is fulfilled by some models of footpoint shuffling and is also satisfied by models with a sufficient level of magnetic resistivity, as outlined above.
Our observations would not be consistent with a cartoon picture of highly braided magnetic field in which bundles of magnetic field lines or flux tubes are wound around each other on length-scales that are large compared with the diameter of each bundle. While still being braided, the field relaxes while driven and would not reach the highly braided state as often depicted in cartoons.

Only if a highly braided state is prescribed
as the initial condition in a 3D model,  braided structures might be visible
\cite[][]{2017ApJ...837..108P}.
Such a signature of braiding has been seen in coronal emission during the first flight of Hi-C \cite[][their Fig.\,2]{2013Natur.493..501C}.
This structure at the edge of a sunspot was part of a low-lying twisted magnetic flux rope \cite[][]{2014ApJ...780..102T}.
At the location where the braiding feature is seen, the flux rope forks (see \sect{S:braid}).  
The braiding of this flux rope might be induced by shearing of the magnetic field \cite[][]{2010ApJ...708..314A}. 
Traditional models of braided coronal loops assume that the braiding is internal to the loop, without forking of the magnetic field into multiple anchors at one or both footpoints (see \sect{S:braid}).
The feature seen by \cite{2013Natur.493..501C}
is a braided structure of case (2) with disparate footpoints (as defined in \sect{S:braid}), and not a good prototype of a single coronal loop bundle of case (1) that we investigate here.

A number of promising processes to energize the solar corona also do not show signatures of braiding. Models that account for the key properties of the chromosphere produce spicules in which thin threads of material are injected into the upper atmosphere \cite[][]{2017Sci...356.1269M}. 
These spicules can be associated with heating of coronal plasma \cite[][]{2017ApJ...845L..18D}. Wave heating might produce strands of coronal emission within a loop that do not appear to be braided  \cite[][]{2019A&A...623A..53K}.

In conclusion, our combination of high-resolution data from Hi-C and IRIS with a 3D MHD model provides an understanding of the cool and warm loops in the cores of active regions.
Because the magnetic field emerges while it gets heated from the footpoints, the cool loops are found just grazing above the chromosphere, while the warmer loops are found higher up at the top of the emerging bundle of magnetic field lines. This causes a spatial offset in the vertical and (in general) also in the horizontal direction between loops showing plasma at (slightly) different temperatures.
While they are part of the same larger emerging structure, the cool and warm loops seen in EUV emission in the cores of active regions are still spatially distinct and show little evidence of twist.
These loops can be heated by stressing the magnetic field in the photosphere, e.g. through field-line braiding, but the relaxation of the magnetic field is efficient enough so that no (clear) signatures of this braiding will be observable.


{\small
\begin{acknowledgements}
\noindent
We acknowledge the constructive comments and patience of an anonymous referee that helped to improve the manuscript.
L.P.C.\ received funding from the European Union's Horizon 2020 research and innovation programme under the Marie Sk{\l}odowska-Curie grant agreement No. 707837.
F.C.\ acknowledges the George Ellery Hale Postdoctoral Fellowship offered by the University of Colorado. 
P.T. acknowledges support by contracts 8100002705 and SP02H1701R from Lockheed-Martin to SAO.
%
%
S.K.T. gratefully acknowledges support by NASA contracts NNG09FA40C (IRIS), and NNM07AA01C (Hinode). 
We acknowledge the High-resolution Coronal Imager (Hi-C 2.1) instrument team for making the second re-flight data available under NASA proposal 17-HTIDS17\_2-003.
MSFC/NASA led the mission with partners including the Smithsonian Astrophysical Observatory, the University of Central Lancashire, and Lockheed Martin Solar and Astrophysics Laboratory.
Hi-C 2.1 was launched out of the White Sands Missile Range on 2018 May 29.
%
%
IRIS is a NASA small explorer mission developed and operated by LMSAL with mission operations executed at NASA Ames Research center and major contributions to downlink communications funded by ESA and the Norwegian Space Centre.
%
%
SDO data are courtesy of NASA/SDO and the AIA and HMI science teams.
%
%
Figure\,\ref{F:model.vert.cut}c is generated using the VAPOR tool (www.vapor.ucar.edu). 
%
%
This research has made use of NASA's Astrophysics Data System.
H.P. and F.C. acknowledge PRACE for awarding us the access to SuperMUC based in Germany at the Leibniz Supercomputing Centre (LRZ).
\end{acknowledgements}
}

\vspace{-1ex}




\clearpage

\appendix

\section{Spatial alignment of Hi-C and IRIS data} \label{sec:align}

The images of the Hi-C 172\,{\AA} band and the IRIS 1400\,{\AA} slit-jaw images cannot be spatially aligned directly.
In active regions and assuming equilibrium conditions, these two channels would show plasma predominantly at temperatures just below 1\,MK and at about 0.1\,MK.
Hence their appearance will be quite different in most places.
Still, using the full-disk data from AIA, we can perform the spatial alignment indirectly.

For this indirect spatial alignment we use the 171\,{\AA} and 1600\,{\AA} channels of AIA.
The choice of the 171\,{\AA} channel is obvious, because it is dominated by the same line of \ion{Fe}{9} and has a similar (but more narrow) temperature response function as the Hi-C 172\,{\AA} band \cite[][]{2019SoPh..294..174R}.
The AIA 1600\,{\AA} band is dominated by the Ly-continuum of \ion{Si}{1} and the \ion{C}{4} doublet at 1548\,{\AA} and 1550\,{\AA}.
The IRIS 1400\,{\AA} band is mainly composed by the same \ion{Si}{1} Ly-continuum and the doublet of \ion{Si}{4} at 1393.76\,{\AA} and 1404.77\,{\AA} (with some contribution also from the \ion{C}{2} doublet near 1335\,{\AA}).
Because \ion{C}{4} and \ion{Si}{4} form at similar temperatures, the AIA 1600\,{\AA} and IRIS 1400\,{\AA} bands show, for the most part, similar features.
Consequently, these two bands have been used frequently for spatial alignment, and our study is no exception.

The alignment between Hi-C 172\,{\AA} and AIA 171\,{\AA} is straight forward.
The same is the case for IRIS 1400\,{\AA} and AIA 1600\,{\AA}.
Both can be achieved easily with standard cross-correlation techniques.
Here we use the publicly available procedure rotalign.pro
(by  R. Molowny, available at \url{http://www.staff.science.uu.nl/~rutte101/rridl/dotlib/rotalign.pro}).
This calculates both the lateral and rotational offsets which are then applied to the data.
The residual offset for the Hi-C vs.\ AIA 171\,{\AA} and IRIS vs.\ AIA 1600\,{\AA} is very small, smaller than 0.1{\arcsec}.
So the problem of indirectly aligning Hi-C with IRIS relies on the alignment of the AIA data in the 171\,{\AA} and the 1600\,{\AA} bands.
The remainder of this section will be devoted to this.

The AIA data cover the full solar disk, which is why one can use the position of the limb for a reliable alignment. 
We select the AIA 171\,{\AA} and 1600\,{\AA} images closest in time to the Hi-C frame we concentrate our analysis on (frame \#\,58 at 19:00:33 UT).
These data headers contain information on the alignment of the AIA data provided by the instrument team.
We then use procedures from the Maps package available in SolarSoft (\url{http://www.lmsal.com/solarsoft/}) to co-register the 171\,{\AA} and 1600\,{\AA} images.
These co-registered images are displayed in \fff{F:full.disk}.

To check the (automated) co-registration of the AIA images, we calculate the position of the limb in the 171\,{\AA} and 1600\,{\AA} bands.
For this, we define ten segments at the limb spread over the whole limb but avoiding an active region at the west limb (see \fff{F:full.disk}).
For each segment we determine the intensity variation in the radial direction across the limb averaged in the azimuthal direction (along the limb).
These cuts are presented in \fff{F:cross.limb.profiles} for both the 171\,{\AA} and 1600\,{\AA} bands.
As expected, the solar limb in the coronal images of AIA 171\,{\AA} is further out than in images of AIA 1600\,{\AA} that mostly show the temperature minimum region at the base of the chromosphere.
Here we define the limb position as the (local) maximum just before the intensity drops above the limb and mark these positions by dots in \fff{F:cross.limb.profiles}.

\begin{figure}
\centerline{
\includegraphics[width=0.5\textwidth]{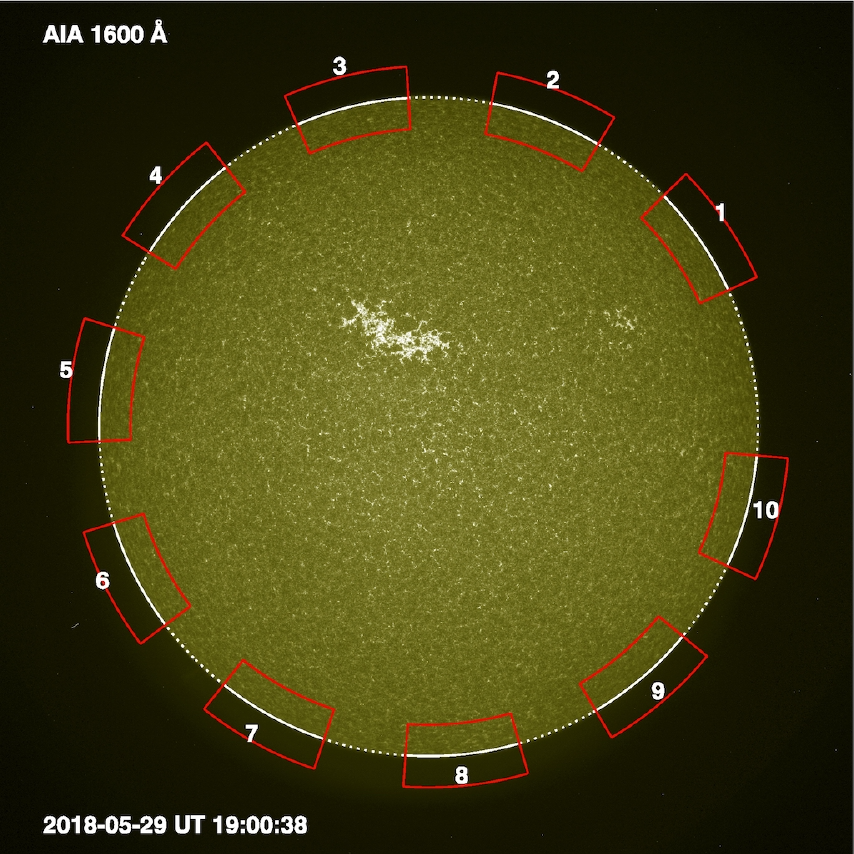}
\includegraphics[width=0.5\textwidth]{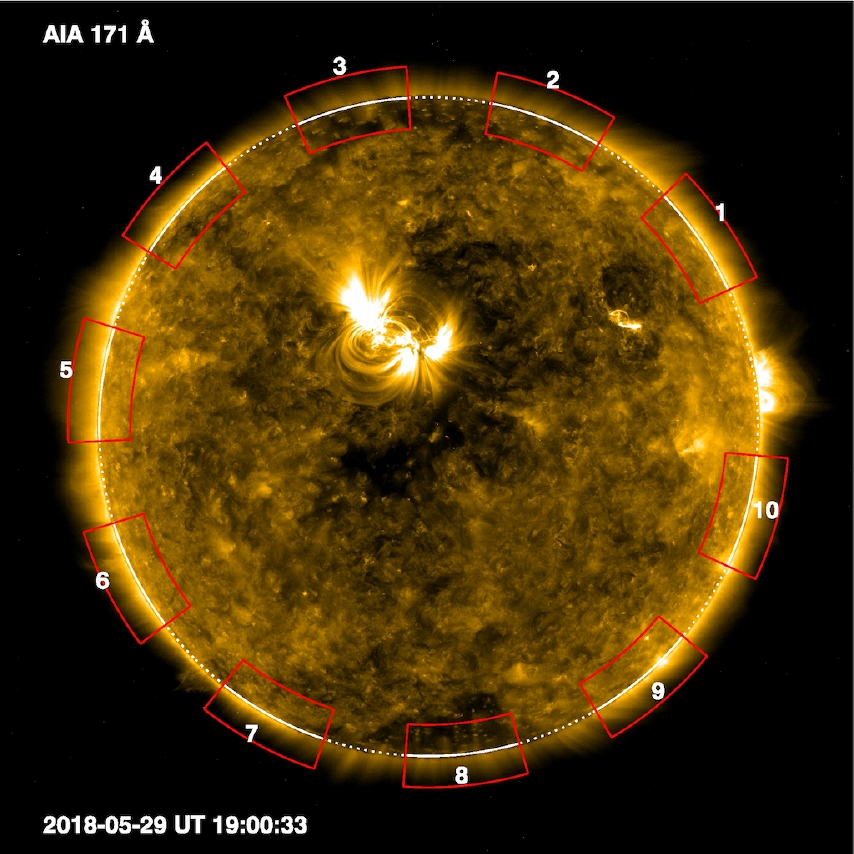}
}
\caption{Full-Sun images acquired by AIA in its 1600\,{\AA} and 171\,channels.
The solid/dotted lines show the limb through a circle with the solar radius as recorded in the header of the AIA 171\,{\AA} image.
The segments numbered 1 to 10 mark the regions to verify the limb alignment between the two channels.
Segments 2, 3, and 8 are at locations where (polar) coronal holes are presents.
The coronal holes are revealed by lower intensities in the 171\,{\AA} image.
The target region of the Hi-C campaign is the bright active region north-east (top-left) of disk center.
See \app{sec:align}. 
\label{F:full.disk}}
\end{figure}

\begin{figure}
\centerline{\includegraphics[width=0.7\textwidth]{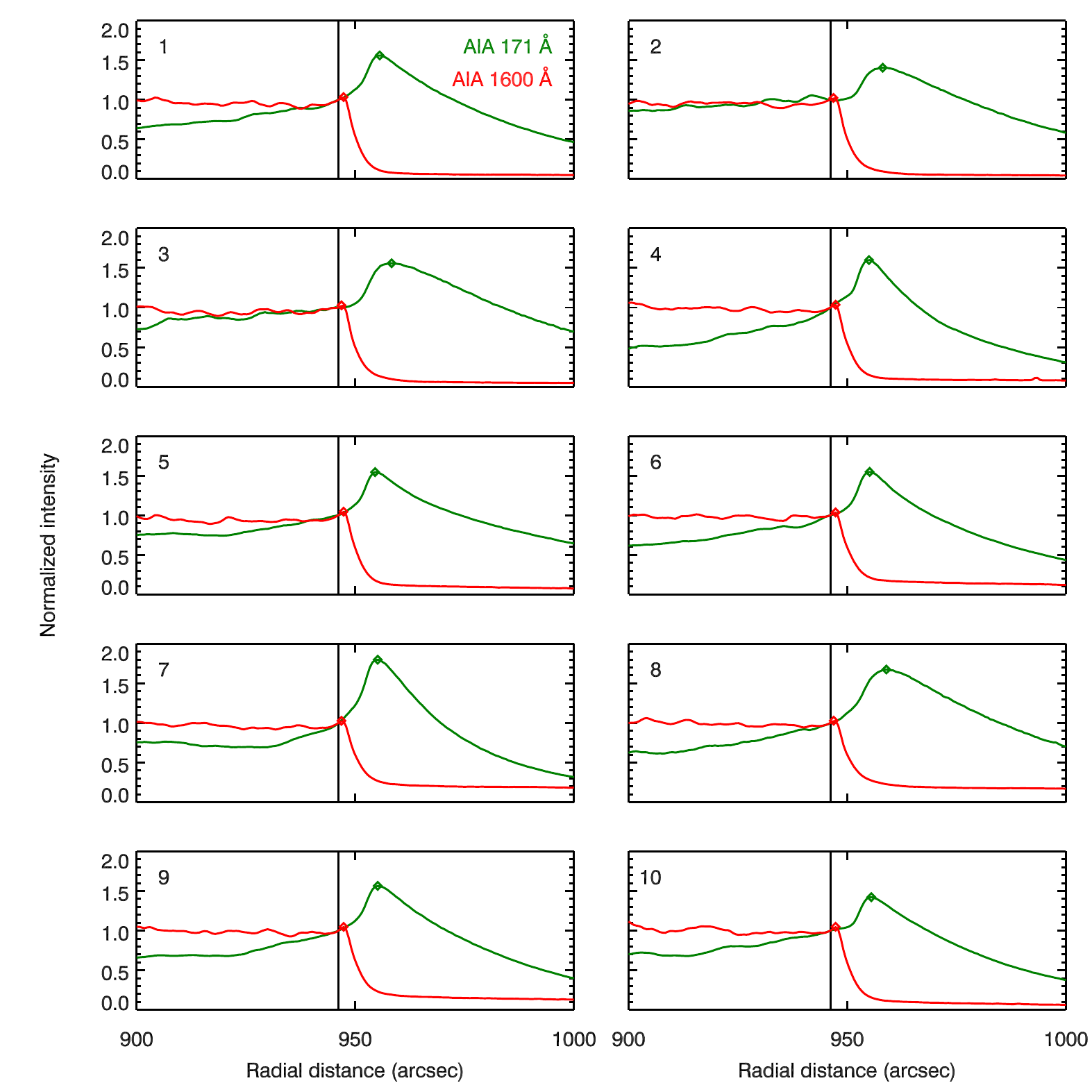}}
\caption{Limb positions in the AIA full-disk data.
We show the intensity as a function of radial distance across the limb averaged in the azimuthal direction (i.e. along the limb) for the ten limb segments marked in \fff{F:full.disk}.
The red lines are for the 1600\,{\AA} channel, the green lines for the 171\,{\AA} band.
The dots mark the respective limb positions defined as the local maximum before the intensity drops above the limb.
The radial distance is measured from disk center.
See \app{sec:align}.
\label{F:cross.limb.profiles}}
\end{figure}

If the AIA 171\,{\AA} and 1600\,{\AA} images are aligned well, then the distance between the limb positions in the two channels should be the same for all the limb segments we choose in \figs{F:full.disk} and \ref{F:cross.limb.profiles}.
These differences between the AIA 171\,{\AA} and 1600\,{\AA} limb positions are shown in \fff{F:limb.pos} for these ten segments.
We can discard segments 2, 3 and 8 because these are at locations of (polar) coronal holes as revealed by their darker appearance in the AIA 171\,\AA\ image in \fff{F:full.disk}.
The intensity as a function of height in a coronal hole is different due to different plasma properties; we discard them to remove this uncertainty. 
The other segments all show offsets around a mean value of 7.88{\arcsec} and a standard deviation of 0.39{\arcsec}.
From this we can conclude that the co-registration of the AIA 171\,{\AA} and 1600\,{\AA} images is good and within about 0.4{\arcsec}.
Together with the good alignment of Hi-C vs.\ AIA 171\,{\AA} and IRIS vs.\ AIA 1600\,{\AA} we estimate that the final alignment between Hi-C and IRIS should be better than 0.5{\arcsec}, i.e.\ slightly better than the AIA pixel size of 0.6{\arcsec}.

To further confirm our (automated) co-registration of the AIA 171\,{\AA} and 1600\,{\AA} images, we align these two images using a recent beta version of the aia$\_$prep routine (at the time of writing the first draft not yet available publicly, G.~Slater, private communication). This beta version co-registers the images by taking into account the limb positions determined from a three-hour running average. This procedure essentially removes any jitter and thermal drifts in the pointing information that could affect the co-registration. With this independent check, we find that all the aforementioned segments (excluding 2, 3 and 8) show offsets around a mean value of 8{\arcsec} and a standard deviation of 0.29{\arcsec}.
This slightly differs (by about 0.1{\arcsec}) from the standard AIA procedures and is well within the errors we estimated for the alignment of the AIA 171\,{\AA} and 1600\,{\AA} images.

\begin{figure}
\centerline{\includegraphics[width=0.5\textwidth]{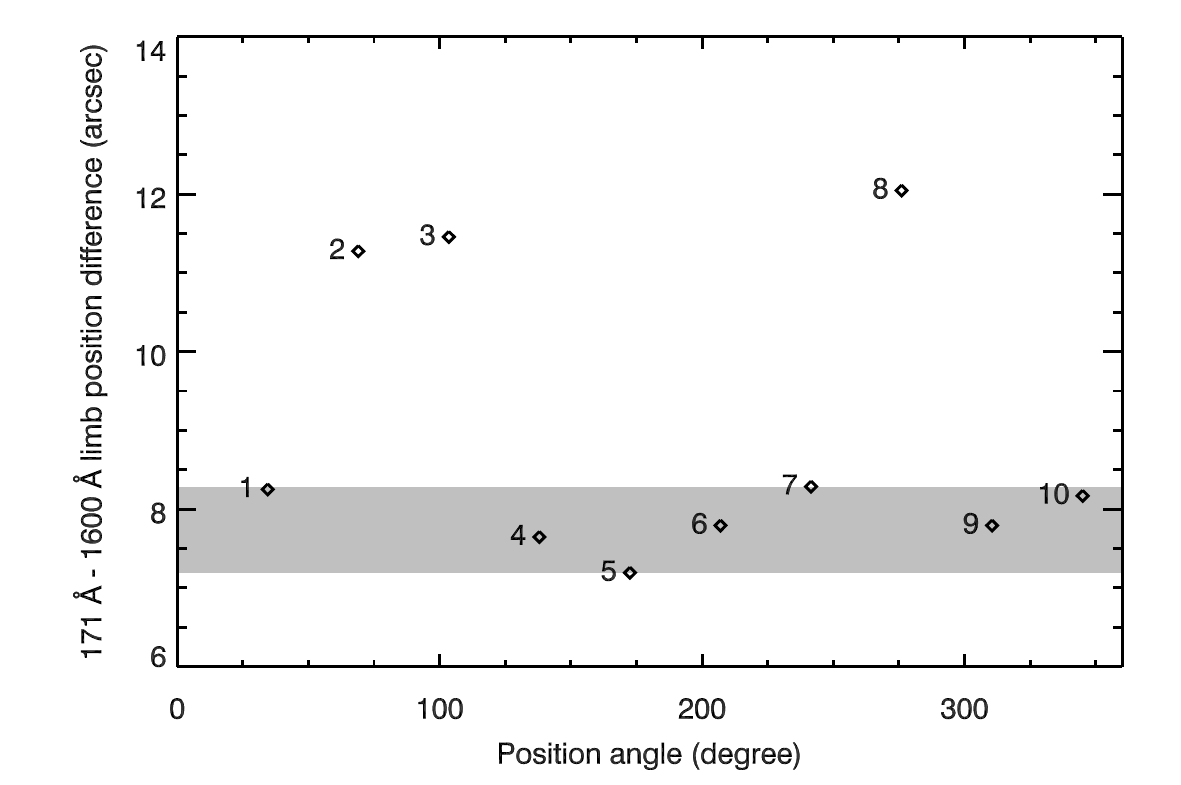}}
\caption{Offset of the limb position between the AIA 171\,{\AA} and 1600\,{\AA} images.
The numbered diamonds show the difference of the limb position between the two bands as derived from the respective intensity variations in \fff{F:cross.limb.profiles} in the segments marked in \fff{F:full.disk}.
The grey band indicates the standard deviation around the mean value (discarding segments 2, 3, and 8 are within the polar coronal holes). 
See \app{sec:align}.
\label{F:limb.pos}}
\end{figure}

\clearpage

\section{Examples of parallel loops in Hi-C and IRIS\label{sec:samples}}

In the main text we discuss the offsets between loops seen in Hi-C and IRIS of two features marked [1] and [2] in \fff{F:hic.core}.
Here we will investigate further examples of such cool and warm loops running parallel to each other in the core of the active region.
We show the zooms of Hi-C image cut-outs and the variation of the intensity across the respective loop feature as seen in Hi-C and IRIS in \fff{fig:ex.loops}.
Each of these further examples is displayed in the same format as the plots for features [1] and [2] in \fff{F:hic.cuts}.
Some of these examples overlap with examples used in a study on the role of magnetic flux cancelation for fine-scale explosive energy release \cite[][]{2019ApJ...887...56T}, based on the same Hi-C data data set.

We determine the offsets between the loops in Hi-C and IRIS by calculating the position of the respective feature along the cross-sectional cuts (roughly) perpendicular to the loop.
We define the location of a loop as the position of the peak intensity along the cut as following from a cubic interpolation.
Of course, this is only one possible choice.
We preferred this over the calculation of the centroid of the cut or a Gaussian fit, because the profiles are asymmetric and have different background levels on both sides of the loop.
We define the spatial offset as the difference in location of the loops in Hi-C and IRIS data, and these offsets are noted with the plots in \fff{fig:ex.loops} and listed in \tab{T:offsets}.
Feature [6] in \fff{fig:ex.loops} illustrates the accuracy we can expect for the determination of the offsets.
While having an offset of just below 0.1{\arcsec}, this offset is clearly visible, even though this is significantly smaller than the spatial resolution of both Hi-C and IRIS (i.e. about 0.3{\arcsec} to 0.5{\arcsec}).
This is because a centroid (or the peak) of a profile can be determined with sub-pixel or sub-resolution-element accuracy as long as the profile is smooth and not noisy.
This visual error estimation is confirmed by a Monte-Carlo-type analysis.
From 300 normal random realizations of photon noise added to both the Hi-C and IRIS maps, we repeated the determination of the offsets between the Hi-C and IRIS loops.
This provides a more rigorous error estimate and results also in a  typical  uncertainty of 0.1{\arcsec}.

One of the thinnest loops we noticed in the active region core is feature [4] in \fff{fig:ex.loops}.
In Hi-C the full width at half maximum is about 0.5{\arcsec}.
So clearly the spatial resolution of Hi-C is at least as good as 0.5{\arcsec}.

The offsets between the loops seen in Hi-C and IRIS range from essentially zero, i.e. no offset, to more  than 0.7{\arcsec} (\tab{T:offsets}).
The largest relative offset between two features is found between [5] and [8] and is more than 1.4\arcsec.
Still, it could be speculated if there is an alignment between Hi-C and IRIS so that most the offsets get close to zero.
To illustrate the offsets graphically, we show these in \fff{fig:offsets} on a plane with the offsets separately in the $x$ and $y$ directions, ${\Delta}x$ and ${\Delta}y$.
Because of the large-scale orientation of the magnetic field in the active region core, the core loops are (an average) generally oriented in a direction from North-East to South-West (see \fff{F:hic.core}b), as indicated by the dashed line in \fff{fig:offsets}. Because we determine the offsets perpendicular to the loops, the data points in \fff{fig:offsets} are mostly found roughly along a line perpendicular to the average loop direction.
Of course, some features show only small offsets, but others have significant offsets, and towards different sides of the respective loop.
So, we can safely conclude that the loops in Hi-C and IRIS cannot all be brought to overlap by applying a shift of the IRIS and Hi-C images with respect to each other.
Instead, in general the loops seen in Hi-C and IRIS have to be at different locations and essentially run in parallel with small but clearly resolvable offsets.

Here we make the implicit assumption that the images in both Hi-C and IRIS are perfectly flat.
In principle, also optical deformations could lead to variable offsets across the field-of-view of both instruments.
Some structures are very close to each other and still show a significant differential offset.
For example, features [2a] and [2b] are less than 2{\arcsec} apart and show a differential offset of 0.2{\arcsec}, i.e.\ two pixels of Hi-C.
Based on still images alone we cannot finally rule out that optic deformations of Hi-C and/or IRIS would cause our offsets.
However, the temporal evolution shows that the features seen in Hi-C and IRIS are at different temperatures (cf.\ \sect{sec:tevol}), which gives us reassurance that the offsets we see are a real effect on the Sun.

\begin{figure}
\centerline{\includegraphics[width=\widcuts]{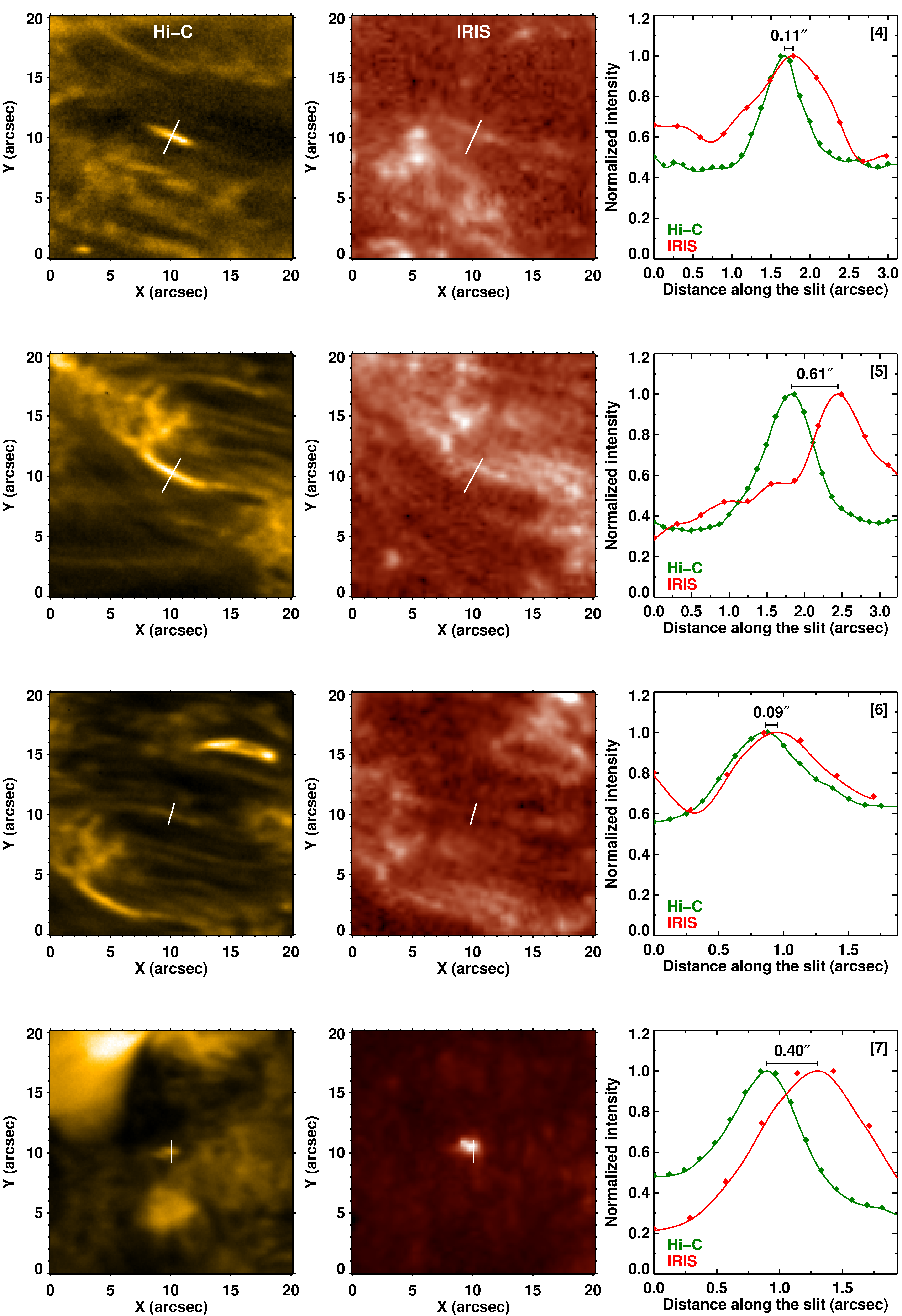}}
\caption{Examples [4] -- [7] of loops with spatial offsets between Hi-C and IRIS. 
The panels have the same format as used for the features in \fff{F:hic.cuts}.
For each feature the left panels shows a zoom into the Hi-C and IRIS images with the white line marking the cut.
The respective right panel shows the intensities in Hi-C (green) and IRIS (red) across the loop along the cut.
The diamonds show the data interpolated on the plate scale of the  respective instrument, the solid curves show interpolations at higher resolution to measure the offsets between the two instruments.
The offsets are indicated with the plots and are listed in \tab{T:offsets}.
The typical error for the offsets is 0.1{\arcsec}.
The distance along the slit is counted positive towards the northern (large values of Y) tip of the cut.
See \app{sec:samples}.
\label{fig:ex.loops}}
\end{figure}

\begin{figure}
\centerline{\includegraphics[width=\widcuts]{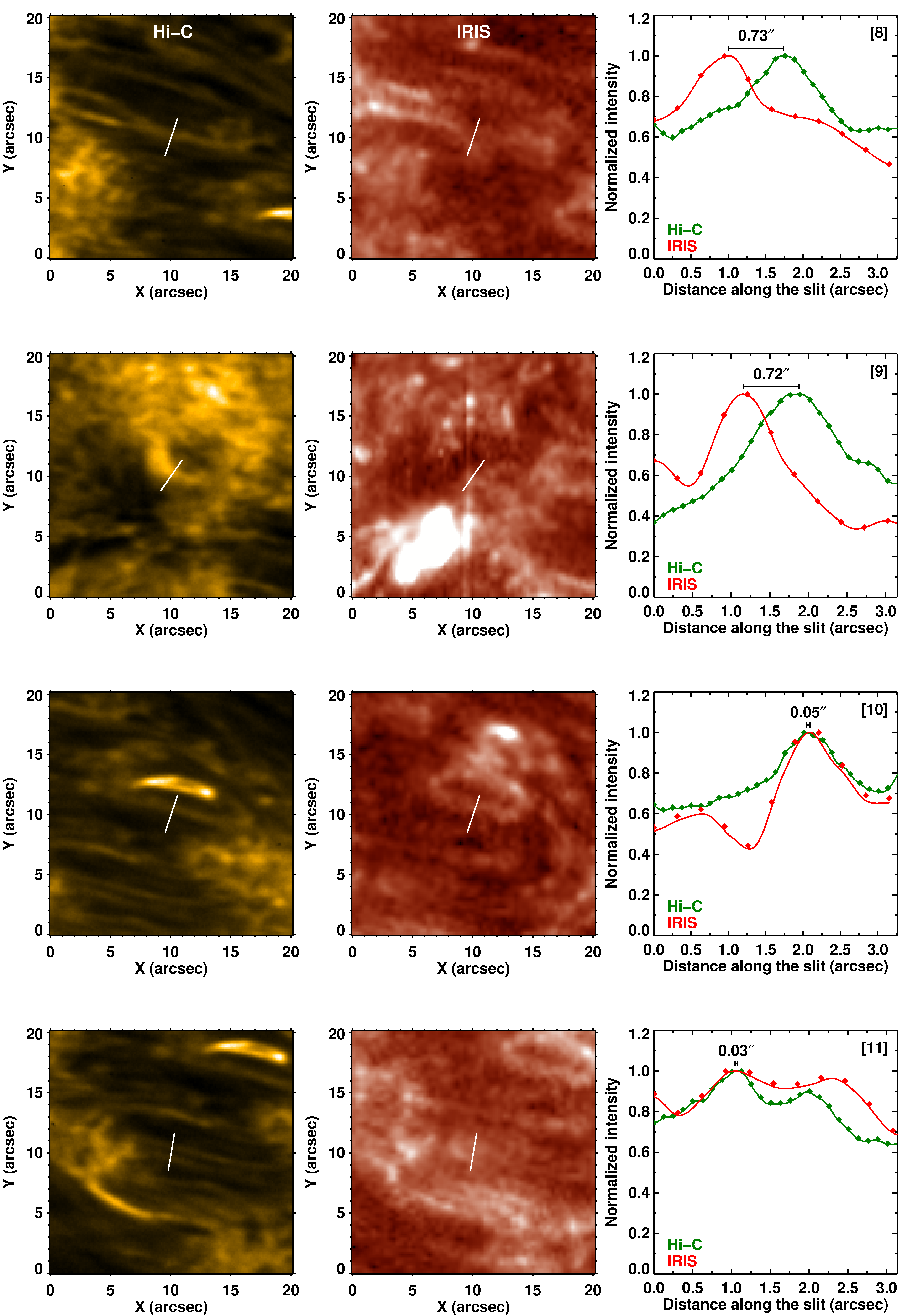}}
\caption{Further examples for offsets. Same as \fff{fig:ex.loops}, but for cases [8--11] in \tab{T:offsets}.
See \app{sec:samples}.
\label{fig:ex.loops.B}}
\end{figure}

\clearpage

\begin{table}
\caption{Offsets between loops seen in Hi-C and IRIS.\label{T:offsets}}
\begin{center}
\begin{tabular}{lcccccccccccccccccccccc}
\hline \hline
feature{$^\dag$}       & [1] & [2a] & [2b] & [4] & [5] & [6] & [7] & [8] & [9] & [10] & [11] \\
$|$offset$|${$^\ddag$}
 &  0.53\arcsec   
 &  0.46\arcsec   
 &  0.28\arcsec   
 &  0.11\arcsec   
 &  0.61\arcsec   
 &  0.09\arcsec   
 &  0.40\arcsec   
 &  0.73\arcsec   
 &  0.72\arcsec   
 &  0.05\arcsec   
 &  0.03\arcsec   
\\ \hline
\end{tabular}
\\[1ex]
{\small
{$^\dag$~}{Features [1] and [2a,b] are shown in \fff{F:hic.cuts}, the other features are displayed in \fff{fig:ex.loops}.}
\\
{$^\ddag$~}{The offsets listed here are the absolute values.
See \fff{fig:offsets} for the offsets in the $x$ and $y$ directions.
Their typical error is 0.1{\arcsec}.
}
}
\end{center}
\end{table}

\begin{figure}
\centerline{\includegraphics[]{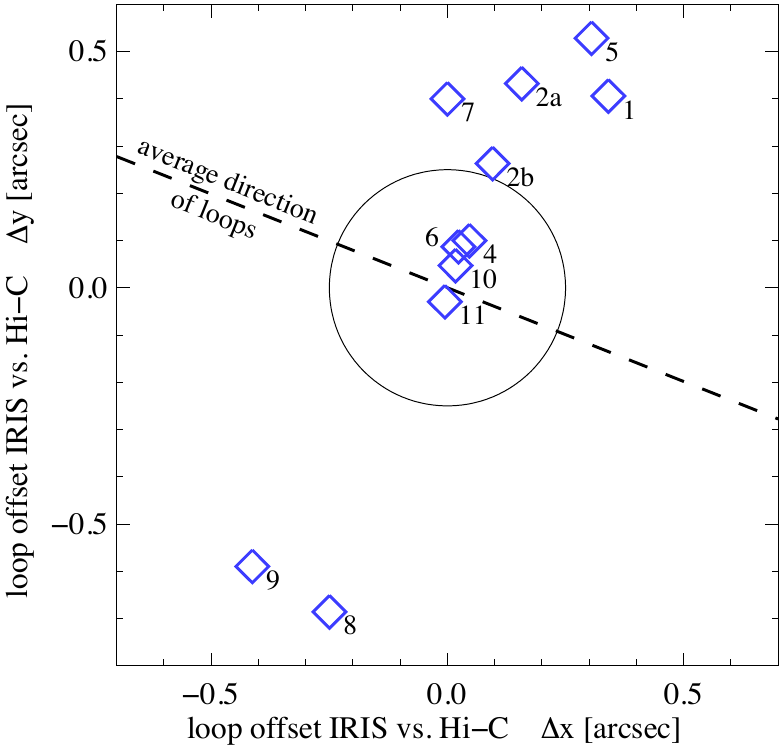}}
\caption{Offsets between loops in Hi-C and IRIS.
Each diamond represents one feature labeled in the same scheme as in \fff{F:hic.cuts}, \fff{fig:ex.loops} and \tab{T:offsets}.
The dashed line shows the average direction of the loops (cf. \fff{F:hic.core}b) reflecting the overall orientation of the magnetic features.
The circle has a diameter of 0.5{\arcsec} that is a slightly larger than the resolution of either Hi-C or IRIS.
See \app{sec:samples}.
\label{fig:offsets}}
\end{figure}

\section{A peculiar short loop \label{S:short}}

While not being the main topic of the study, we want to use this opportunity to highlight a peculiar short loop labeled as feature [3] in \fff{F:hic.core}.
The main goal of this short discussion is to report a novel feature that might or might not be found regularly on the Sun.
This feature has been discussed also in the study on flux emergence and explosive energy release based on the very same Hi-C data set \cite[][]{2019ApJ...887...56T}, but not in the following sense of temporal evolution and heating.

In the small rectangle in \fff{F:hic.core}b, a short loop feature in the Hi-C image is visible with with a length of less than 5\,Mm. This small feature even seems to show some sub-structure.
Also in the first flight of Hi-C, such short miniature loops were seen in coronal emission \cite[][]{2013A&A...556A.104P} --- those were even shorter than our example here.
In contrast to the examples from the first Hi-C flight, we can clearly identify the loop footpoints at opposite magnetic polarities (\fff{F:hic.core}c).

Most interestingly, this loop seems to be surrounded by emission from cooler plasma as seen in the IRIS 1400\,{\AA} slit-jaw images, as it is evident from the multi-color composite image in \fff{F:hic.core}d.
There it appears as if the warm loop in Hi-C has a cool halo in IRIS. Actually, at the place of the short loop in Hi-C, there is a hole in the emission of IRIS 1400\,{\AA} (\fff{F:hic.core}a). %
This indicates that the core of a cooler structure has been heated to higher temperature, with the plasma in the outer part being not (or less) heated. However, the short time interval covered through the suborbital rocket flight prevents a more detailed analysis of the temporal evolution.

\clearpage

\section{Temperatures of the plasma seen by H{\lowercase{i}}-C, AIA, and IRIS} \label{S:temp.diag}

In \sect{S:temp.IRIS.AIA} we discuss the possible temperature(s) in the source regions of the emission recorded by Hi-C, AIA and IRIS. The main conclusion is that Hi-C sees plasma at higher temperatures than IRIS does. Here we follow a procedure outlined by \cite[][]{2013ApJ...771...21W} to check if the source region of Hi-C (or AIA) is closer to 0.3\,MK or 0.8\,MK by investigating the temporal variation of the intensity in the various channels.

\subsection{DEM and EM-loci methods}

Usually, one would employ an inversion and obtain the differential emission measure (DEM) to get a measure at which temperature(s) the bulk part of the plasma is radiating \cite[e.g.][]{2015ApJ...807..143C}. 
However, when using AIA data, these inversions are not reliable for temperatures below about ${\log}T\,[{\rm{K}}]{\approx}5.7$, because of the limited temperature sensitivity in the AIA channels for those low temperatures.
Furthermore, one has to treat these inversions with care if applied to small dynamic structures, and the features we investigate here change on time scales of one minute (e.g.\ \figs{F:hic.time} and \ref{fig:ex.tevol}). This is fast, because it is comparable to or smaller than the radiative cooling time. The latter can be defined as the time scale to radiate the internal energy. Using the radiative loss function from Chianti
\cite[][]{2019ApJS..241...22D}, for a density of a low-lying loop of $10^{10}$\,cm$^{-3}$ and temperatures from 0.3\,MK to 1\,MK this ranges from 20\,s to 60\,s. Performing a sparse DEM analysis \cite[][]{2015ApJ...807..143C} we typically find peaks of the DEM around $\log{T\,[{\rm{K}}]}{\approx}5.8$ and $6.2$ in the  cool and warm loops in the active region core. However, we get similar results if we check a neighboring quiet region or a moss area. So we cannot find a reliable temperature estimate through the DEM inversion.   

An alternative to the DEM analysis is the emission-measure (EM) loci method. Under isothermal conditions, the EM curves from different channels of AIA would intersect at one temperature that characterizes the plasma emission.
This technique has been applied, e.g., to inter-moss loops that show simultaneous brightening in extreme-UV channels and provided evidence that those loops are at temperatures significantly below 1\,MK, at ${\approx}$0.3\,MK \cite[][]{2013ApJ...771...21W}.
However, for our data set this technique suffers from the same problems as the DEM inversion. 
Furthermore, when we apply this technique to the model results (see \sect{S:model}) it also returns temperatures around 0.3\,MK, while in the model we find the true temperature distribution in the loop to be at significantly higher values, closer to 1\,MK.

\subsection{Temporal evolution and implications for loop temperature\label{S:temp.evol.contrib}}

Acknowledging the limitations of the DEM and EM-loci techniques, we restrict our analysis to a direct look at the light curves from the loops and relate this to the temperature contributions of the respective channels.
In \fff{fig:contrib.fct} we show the contributions (sometimes refereed to as temperature kernels) for the AIA channels at 131\,{\AA}, 171\,{\AA}, 193\,{\AA}, and 211\,{\AA}.
The main peaks of their contributions range from 0.6\,MK to 2\,MK.
However, except for the 131\,\AA\ channel, the other three have a clear secondary peak at about 0.3\,MK (highlighted in\ \fff{fig:contrib.fct}), originating from transition region lines such as \ion{O}{5}.

Following the temporal evolution of the loops, i.e.\ their light curves, one might be able to distinguish at least if the contribution in the AIA channels would be around 0.3\,MK or at significantly higher temperatures.
If the light curves of the AIA channels are evolving in the same way, there are two possible interpretations for the observed loop:
(1) The loop is multi-thermal, i.e.\ it consists of a mixture of (non-resolved) strands at different temperatures that are heated simultaneously, or 
(2) the loop has a temperature well below 1\,MK, probably around 0.3\,MK, as argued for the inter-moss loops \cite[][their Sect.\ 2.2]{2013ApJ...771...21W}.

The fast evolution seen in the light curves raises the question of the validity of ionisation equilibrium that is used to calculate the contribution functions in \fff{fig:contrib.fct}.
Considering their length, the loops we investigate here most probably will be low-lying compact objects in the active region core. 
If they indeed do not reach too high temperatures and assuming a constant pressure, they will be at comparably high density. This would shorten ionisation and recombination times. 
In their numerical model, \cite{2006ApJ...638.1086P} investigated the ionisation and recombination times for \ion{O}{5}, that will be a major contributor to the side peaks around 0.3\,MK of AIA bands (cf.\ \fff{fig:contrib.fct}).
For \ion{O}{5} they find these time scales to be only 30\,s to 40\,s on average \cite[Fig.\,4 of][]{2006ApJ...638.1086P}. Consequently, the assumption of ionisation equilibrium might be reasonable here.

\subsection{Temporal evolution of loops in different bands\label{S:res.temp.detail}}

We now turn to the application of the temporal evolution of the features in different wavelength bands to get an estimate of their temperature.
The light curves for selected loops for the AIA channels is displayed in \fff{fig:ex.tevol}.
We also add the light curves for the loops as seen in the IRIS 1400\,{\AA} channel, so that the panels in \fff{fig:ex.tevol} are extended versions of \fff{F:hic.time}.
Here we do show a normalization for each of the light curves that maps the minimum and maximum values during the time considered to the range $[0,1]$. %
This will highlight small changes above a (slowly varying) background. 
We do not show the Hi-C 172\,\AA\ light curves, because as expected they are almost exactly the same as for the AIA 171\,{\AA} band.

In principle we could perform this analysis for all the loop samples discussed in \app{sec:samples}, because the data sets from AIA and Hi-C are properly aligned.
However, the spatial resolution of AIA is not sufficient to pick up the other loops properly and would most probably only show the variability of the background.
Consequently we refrain from analysing these other cases.

Of the cases we show, feature [5] is closest in temporal evolution when compared to inter-moss loops \cite[][]{2013ApJ...771...21W}.
Here all the AIA channels evolve very similarly with a transient brightening lasting for about 2 minutes (see \fff{fig:ex.tevol} [5]).
From this we would conclude that the loop seen in 171\,\AA\ (and in the other AIA channels) is at a temperature of around 0.3\,MK. 
Still, the temporal variation in the IRIS 1400\,\AA\ channel (dominated by \ion{Si}{4}) is significantly different in that it increases much slower than the AIA channels. Comparing the contribution function of \ion{Si}{4} in \fff{fig:contrib.fct}, it is clear that most probably the emission we see in IRIS is originating from plasma  that is cooler (${\approx}0.1$\,MK) than the emission from the 171\,\AA\ channel (and hence Hi-C\,172\,\AA), the latter originating from 0.3\,MK or above.

 The other examples show a less clear picture. Feature [2a] shows a two-peaked structure in the AIA channels, with one peak around 18:57 and another around 19:00\,UT (see \fff{fig:ex.tevol} [2a]). However, during the first peak the brightening is strongest in 211\,\AA, while during the second one 131\,\AA\ and 171\,\AA\ show the strongest intensity increase. Thus we might consider in this case the AIA emission to originate not from 0.3\,MK, but from a different (higher) temperature, probably above 1\,MK during the first and 0.6 to 1\,MK during the second peak. As mentioned already with \fff{F:hic.time}, here the peaks in the 1400\,\AA\ channel of IRIS (i.e.\ \ion{Si}{4}, $<$0.1\,MK) are not coinciding in time with the peaks in the AIA channels, so also in this case the structures seen in IRIS and Hi-C\,172\,\AA\ will be at different temperatures and thus they will be different loops.

Only in one case, in feature [4], we see a very similar variation between the IRIS 1400\,\AA\ light curve and the AIA channels (see \fff{fig:ex.tevol} [4]).
So in this case the emission in IRIS and Hi-C might actually originate from the same short loop at a low temperature between 0.1\,MK and 0.2\,MK. Checking the offsets between the IRIS and Hi-C loops in \tab{T:offsets}, reveals that this is one of the three cases with offsets around 0.1\arcsec\ or below. This is within the limits of the alignment procedure (about 0.1\arcsec; see \app{sec:samples}). So it might be that in this case the Hi-C and IRIS loops are indeed co-spatial and at the same temperature.
Feature [7] is similar
to Feature [4] in that there the AIA light curves are similar to each other (though to a lesser extent).   Also the IRIS 1400\,\AA\ variation
(see \fff{fig:ex.tevol}
[7]) is related: while a first peak around 18:58\,UT is similar between IRIS and AIA,
the second peak just before 19:00 is broader in IRIS. So the similarity between IRIS and AIA is less clear than in feature [4], but still present.

\end{document}